\documentstyle[psfig]{mn}

\title[Dwarf Nova Oscillations and Quasi-Periodic Oscillations in Cataclysmic Variables: IV.]
{Dwarf Nova Oscillations and Quasi-Periodic Oscillations in Cataclysmic Variables, IV. Observations 
of Frequency Doubling and Tripling in VW Hyi.}
\author[Brian Warner and Patrick A. Woudt]
       {Brian Warner\thanks{email: warner@physci.uct.ac.za} and 
        Patrick A. Woudt\thanks{email: pwoudt@circinus.ast.uct.ac.za}\\
        Department of Astronomy, University of Cape Town, Private Bag,
        Rondebosch 7700, South Africa}
\date{2005 March 25}

\begin{document}

\maketitle

\begin{abstract}
We present new observations of the rapid oscillations in the dwarf nova VW Hyi, 
made late in outburst. These dwarf nova oscillations (DNOs) increase in period until they 
reach 33 s, when a transition to a strong 1st harmonic and weak fundamental takes place. After 
further period increase the 2nd harmonic appears; often all three components are present simultaneously. 
This 1:2:3 frequency suite is similar to what has been seen in some neutron star and black hole X-Ray 
binaries, but has not previously been seen in a cataclysmic variable. When studied in detail the 
fundamental and 2nd harmonic vary similarly in phase, but the 1st harmonic behaves 
independently, though keeping close to twice the frequency of the fundamental.
The fundamental period of the DNOs, as directly observed or inferred from the harmonics, increases 
to $\sim$ 100 s before the oscillation disappears as the star reaches quiescence. Its maximum period 
is close to that of the `longer period' DNOs observed in VW Hyi.
The quasi-periodic oscillations (QPOs), which have fundamental periods 
400 -- 1000 s, behave in the same way, showing 1st and 2nd harmonics at approximately the same 
times as the DNOs.
We explore some possible models. One in which the existence of the 1st harmonic is due to transition 
from viewing a single accretion region to viewing two regions, and the rate of accretion onto the 
primary is modulated at the frequency of the 1st harmonic, as in the `beat frequency model', can 
generate the suite of DNO frequencies observed. But the behaviour of the QPOs is not yet understood.
\end{abstract}


\section{Introduction}

      Many dwarf novae show rapid (typical periods 5 -- 40 s) low amplitude brightness 
oscillations during outburst, and some nova-like variables show similar modulations. These 
have been actively studied since their discovery some thirty years ago (Warner \& Robinson 1972) and 
are known collectively as dwarf nova oscillations (DNOs). They show a systematic period-luminosity 
relationship during outburst, with minimum period at maximum bolometric luminosity, and are of 
moderate coherence near maximum but usually appear to become increasingly incoherent near the 
return to quiescence. The DNO periods have been observed to range over a factor of more than 
two in some dwarf novae. The short periods indicate that the source of these modulations is 
close to the white dwarf primary in these cataclysmic variables (CVs); this is confirmed by 
the appearance of oscillations of substantial amplitude in soft X-Ray and EUV flux during 
the outbursts of, e.g., the nearby dwarf novae VW Hyi and SS Cyg. Evidently the mass transfer 
from the accretion disc to the primary is strongly modulated at the DNO period.

    The DNOs were reviewed in Warner (1995a); updates are given in Warner (1995b), in the 
first two papers in this series (Woudt \& Warner 2002: hereafter Paper I; Warner \& Woudt 2002: 
hereafter Paper II) and in Warner (2004). A model in which the accretion flow close 
to the primary is controlled by the magnetic field generated in a rapidly rotating 
equatorial belt, formed as a result of the high rate of accretion, has been proposed 
(Warner 1995b; Paper II); this is basically an intermediate polar structure, including 
a truncated accretion disc, but where only the belt receives the accreting angular momentum; 
the belt is magnetically coupled to the accretion disc but is able to slip relative to the body 
of the white dwarf primary. This Low Inertia Magnetic Accretor (LIMA) model (which was first 
proposed by Paczynski (1978)) allows large variations in DNO periods, and also explains 
the rapid deceleration of the DNOs observed in VW Hyi near the end of its outburst 
(Papers I and II) -- from the effect of outward propellering of gas as the belt continues to 
revolve rapidly after the mass transfer rate ($\dot{M}$) diminishes. It is probable 
that the inner edge of the truncated disc is close to the corotation radius (where 
the Keplerian period in the disc equals the rotation period of the primary), even 
when propellering is present, as argued by Rappaport, Fregeau \& Spruit (2004). 
Furthermore, the rotation period of the field carried round by the equatorial belt 
will rarely be exactly the same as the period of the inner edge of the disc -- the belt 
is continually being spun up or down by its magnetic connection to the disc. 
This corresponds to the model with large $\Theta$ in the 3D simulations of accretion 
from a disc onto an inclined dipole by Romanova et al.~(2004), which is the model 
in which they find rapid quasi-periodic luminosity variations. In these 3D models, unlike
the simple 2D LIMA model, the field inflates, opens out and reconnects, but the process is
still driven by differential rotation between star and disc (see Uzdensky (2004) for a review).

   The optical DNOs are the result of soft X-Ray and EUV emission from the accretion 
zones creating a `beam' that sweeps around and is reprocessed to longer wavelengths 
by the surface of the disc or anything else it illuminates. Another DNO component 
can come from the accretion zone itself.

    Independent deduction of the existence of a spinning belt in VW Hyi and in other 
dwarf novae has been obtained from HST and FUSE spectroscopic observations. Spectra taken 
after outburst can only be modeled by composite spectra, in the case of VW Hyi comprising a 
white dwarf with projected rotational velocity $v \sin i$ in the range 300 -- 400 km s$^{-1}$ 
and a hotter component of small area and $v \sin i$ 3000 -- 4000 km s$^{-1}$ 
(Sion et al.~1996; G\"ansicke \& Beuermann 1996; Godon et al.~2004; Sion et al.~2004a,b,c). 
The spinning accretion belt lasts for at least 14 days after a superoutburst (Sion et al.~2004b).

    There is a second kind of DNO with periods about 4 times those of the standard DNOs, 
with little or no dependence on accretion luminosity. These are termed longer period 
DNOs: lpDNOs (Warner, Woudt \& Pretorius: hereafter Paper III; Pretorius 2004).

   In addition to the two types of DNO, there are large amplitude modulations with 
periods about 15 times larger than those of the DNOs, and much shorter coherence, 
known as quasi-periodic oscillations (QPOs). In Paper II we proposed that these are 
caused by traveling waves excited in the inner disc -- reprocessing and obscuring 
radiation from the bright central regions of the disc. Furthermore, the occasional 
appearance of `double DNOs' with frequency difference equal to the QPO frequency are 
attributed to the DNO `beam' sweeping around and being partly reprocessed by the 
progradely traveling vertically thickened disc. We should therefore in general be 
prepared to distinguish between `direct' DNOs, which arise from reprocessing from the 
disc, and what we will call here `synodic' DNOs, which are reprocessed from some moving 
site (this is discussed in more detail in Section 3). Whereas the direct DNOs 
are observed to be sinusoidal, the synodic DNOs commonly possess significant amplitude 
at their first harmonic. Because the synodic DNOs are observed to be reprocessed from 
a progradely moving source we do not consider a magnetically warped and precessing inner 
disc, which moves retrogradely (Pfeiffer \& Lai 2004).

     In VW Hyi we observed evolution of both the DNO and the QPO periods but the ratio 
$R = P_{QPO}/P_{DNO}$ remained almost constant at $\sim$ 15 (Paper I). It was pointed 
out that a similar ratio is seen in the high and low frequency QPOs observed in X-Ray binaries 
(Beloni, Psaltis \& van der Klis 2002), and that in fact (identifying DNOs and QPOs in CVs as 
the equivalent of the two kinds of QPOs in X-Ray binaries) the behaviour of VW Hyi forms 
an extension two orders of magnitude lower in frequency of the two-QPO correlation seen 
in the latter. An extension to even lower frequencies in other CVs has since been observed (Mauche 2002; Paper III).

    The importance of further observations of DNOs and QPOs in CVs, where they are more 
easily analysed than in the X-Ray binaries, is evident. The rich set of phenomena exhibited 
by VW Hyi (Papers I and II) makes this a prime target for further study. In Section 2 
we report further high speed photometric measurements of VW Hyi made towards the end of 
outburst, supplementing those already presented in Paper I, and revealing the occurrence 
of harmonic 1:2:3 ratios of periods late in outburst. In Section 3 we consider possible 
interpretations in terms of magnetically controlled accretion; in Section 4 we compare our 
VW Hyi results with those seen in X-Ray binaries and 
in Section 5 we briefly summarize the current state of play.

\begin{table*}
 \centering
  \caption{An overview of a section of our data archive of VW Hyi. Observations in the late stages of decline from outburst with DNOs present.}
  \begin{tabular}{@{}lrcrclrrccccc@{}}
   Run        & Tel. & Filter & HJD Start  & $t_{in}$ & Outburst & $T = 0^*$  & Length  & Range $T$  & \multicolumn{3}{c}{DNOs$^{**}$} & QPO  \\
              &      &        & +244\,0000 & (sec)     &Type$^\dag$     & (days) & (hours) & (days) &     &     &  &     \\[10pt]
 S0018 & 20-in & -- &  1572.49752 & 2 &  Normal (L) &  1572.75 &  2.04 &--0.25 $\rightarrow$ --0.17& F &  & &  -- \\
 S6133 & 74-in & I  & 11785.54592 & 2 & Normal (L) & 11785.75 &  2.84 &--0.20 $\rightarrow$ --0.09 & F &  & &  -- \\
 S6184 & 40-in & -- & 11957.27611 & 5  &Normal (L) & 11957.25 &  1.68 & 0.03 $\rightarrow$ 0.10   & F &   & & $\surd$ \\
 S6059 & 40-in & -- & 11580.27697 & 4 & Normal (M) & 11580.20 &  5.27 & 0.08 $\rightarrow$ 0.30   & F &   & & $\surd$ \\
 S0127 & 40-in & -- &  1677.29328 & 4 & Super      &  1677.20 &  3.77 & 0.09 $\rightarrow$ 0.25   & F &   & & $\surd$ \\
 S7621 & 40-in & -- & 13463.22526 & 6 & Normal (L) & 13463.10 &  4.03 & 0.13 $\rightarrow$ 0.34   & F & 1 & &      \\
 S2915 & 40-in & -- &  4937.28242 & 4 & Normal (L) &  4937.10 &  0.73 & 0.18 $\rightarrow$ 0.21   &   & 1 & &  -- \\
 S1307 & 40-in & -- &  2354.38968 & 4 & Normal (M) &  2354.20 &  1.33 & 0.19 $\rightarrow$ 0.25   &   & 1 & & $\surd$  \\
 S7311 & 74-in & B  & 13139.19081 & 4 & Normal (M) & 13138.95 &  5.41 & 0.24 $\rightarrow$ 0.47   &   & 1 & 2 & $\surd$ \\
 S7342 & 40-in & -- & 13155.19077 & 5 & Normal (M) & 13154.90 &  3.07 & 0.29 $\rightarrow$ 0.42   & F & 1 & 2 & $\surd$ \\
 S6316 & 40-in & -- & 12354.23650 & 5 & Normal (L) & 12353.95 &  5.72 & 0.29 $\rightarrow$ 0.53   &   & 1 &   & $\surd$ \\
 S6138 & 40-in & -- & 11898.27800 &4,5& Normal (M) & 11897.85 &  7.63 & 0.43 $\rightarrow$ 0.75   & F & 1 &   & $\surd$ \\
 S7384 & 74-in & -- & 13234.64681 & 4  &Normal (S) & 13234.20 &  0.78 & 0.45 $\rightarrow$ 0.48   &   & 1 &   &  --  \\
 S3416 & 40-in & -- &  5967.55248 & 2 & Normal (M) &  5967.00 &  1.19 & 0.55 $\rightarrow$ 0.60   &   & 1 &   & $\surd$ \\
 S6528 & 74-in & -- & 12520.21754 & 4 & Normal (L) & 12519.65 & 10.20 & 0.57 $\rightarrow$ 0.99   &   & 1 & 2 &  --  \\
 S5248 & 40-in & -- &  8202.36304 &4,5& Normal (M) &  8201.75 &  4.66 & 0.61 $\rightarrow$ 0.81   &   & 1 &   &  --  \\
 S0019 & 20-in & -- &  1573.44068 & 5 & Normal (L) &  1572.75 &  4.30 & 0.69 $\rightarrow$ 0.87   &   & 1 & 2 & $\surd$ \\
 S2623 & 40-in & -- &  3515.30484 &4,5& Normal (M) &  3514.60 &  3.60 & 0.70 $\rightarrow$ 0.85   &   & 1 & 2 & $\surd$ \\
 S7343 & 40-in & -- & 13155.65679 & 5 & Normal (M) & 13154.90 &  0.21 & 0.76 $\rightarrow$ 0.77   & F &   & 2 &  --  \\
 S0484 & 40-in & -- &  2023.30835 & 4 & Super      &  2022.55 &  3.91 & 0.76 $\rightarrow$ 0.92   & F & 1 &   & $\surd$ \\
 S7222 & 74-in & B  & 13007.27570 & 4 & Super      & 13006.40 &  7.54 & 0.88 $\rightarrow$ 1.19   &   & 1 & 2 & $\surd$ \\
 S0129 & 40-in & -- &  1691.29620 & 5 & Super      &  1690.20 &  1.86 & 1.10 $\rightarrow$ 1.18   & F &   &   &  --  \\
 S7368 & 74-in & -- & 13169.48671 & 4 & Normal (M) & 13168.35 &  2.42 & 1.14 $\rightarrow$ 1.24   & F &   & 2 & $\surd$ \\
 S7301 & 40-in & -- & 13087.24153 & 5 & Normal (M) & 13086.10 &  2.94 & 1.14 $\rightarrow$ 1.26   &   &   & 2 & $\surd$ \\
 S1322 & 40-in & -- &  2355.38508 & 4 & Normal (M) &  2354.20 &  4.91 & 1.19 $\rightarrow$ 1.39   &   &   & 2 & $\surd$ \\[5pt]
\end{tabular}
{\footnotesize 
\newline 
$^*$ The time $T = 0$ is based on a template light curve of VW Hyi as defined in Paper I (see figure 5 of Paper I).\\
$^{**}$ F: Fundamental, 1: First harmonic, 2: Second harmonic. $^{\dag}$ S = short, M = medium, L = long.\hfill \\
}
\label{dno4tab1}
\end{table*}

\section{Photometric observations of VW Hyi}

The VW Hyi data presented here are a combination of some archival data (Paper I) and a large
number of new observations. The latter have been made selectively towards the 
ends of outbursts, with the specific aim of studying the DNO and QPO behaviours 
just as the star returns to quiescence. It is in this phase of outburst that 
apparent frequency doubling has been observed, but the evolution from fundamental 
to first harmonic seemed unstructured, even chaotic, and could not be followed with 
the limited observational material available (Paper I). All new observations were 
made with the UCT CCD Photometer (O'Donoghue 1995) operating in frame-transfer mode 
(resulting in zero dead-time and therefore maximum time-resolution) on the 74-in or 40-in reflectors 
at the Sutherland site of the South African Astronomical Observatory. Fortunately 
VW Hyi is circumpolar at Sutherland, which provides an extended observing season.

     In Table~\ref{dno4tab1} we give the log of the high speed photometric 
runs on VW Hyi that are analysed in this paper. These include a few of the runs 
already studied in Paper I. Integration times (Column 4 in Table~\ref{dno4tab1}) range between 2 and 6 seconds.
In general, observations were taken in white light (no filter, see Column 3 in Table~\ref{dno4tab1}). On a few
occasions, a filter was used to avoid saturating the UCT CCD when observing with the 74-in telescope.
The UCT CCD data are subjected to a standard data reduction routine (including flat fielding using
sky flat fields) which outputs aperture and profile-fitted photometry using the
DoPhot routine (Schechter, Mateo \& Saha 1993). For a bright star such as VW Hyi, 
aperture photometry results in the highest signal-to-noise light curve.
The lack of a suitably bright reference star near VW Hyi in the (small) field of view of the
UCT CCD -- $50'' \times 34''$ on the 74-in, and $109'' \times 74''$ on the 40-in reflector, respectively --
means that all data presented here were obtained in clear conditions.

\subsection{The Behaviour of the DNOs}

\subsubsection{The period evolution and appearance of harmonics}

      We have made Fourier transforms (FTs) of the VW Hyi light curves, in general 
following the progress of DNOs by dividing individual runs into subsections. These subsections
range in length between $\sim$ 5 minutes to $\sim$ 1 hour (see Column 3 of Table~\ref{dno4tab2}). 
Within each subsection, we determined the DNO period, its formal uncertainty, and the amplitude 
of the modulation by a non-linear least-squares fit to the data. Even though the DNOs are generally
not directly visible in the light curves, the signal in the FT is obvious and persistent
and often marks the most prominent peak in the FT at high contrast with respect to the noise level (see Fig.~\ref{dno4fig2}).
The identification of further harmonic signals (at exact multiple frequencies) can subsequently be done at
greater sensitivity. Note, however, that we will only detect DNOs that are moderately coherent.
The quantitative results are given in Table~\ref{dno4tab2}, but it is important to note 
that because the DNO period and amplitude are generally changing, the FT detects 
only the sections of light curve that have relatively strong signals -- we will 
later use amplitude/phase diagrams to produce greater sensitivity. 
     Table~\ref{dno4tab2} contains a considerable amount of data. As we are dealing
here with a previously unobserved phenomenon in dwarf novae we consider it important
to provide the evidence and uncertainties in some detail.

\begin{figure}
\centerline{\hbox{\psfig{figure=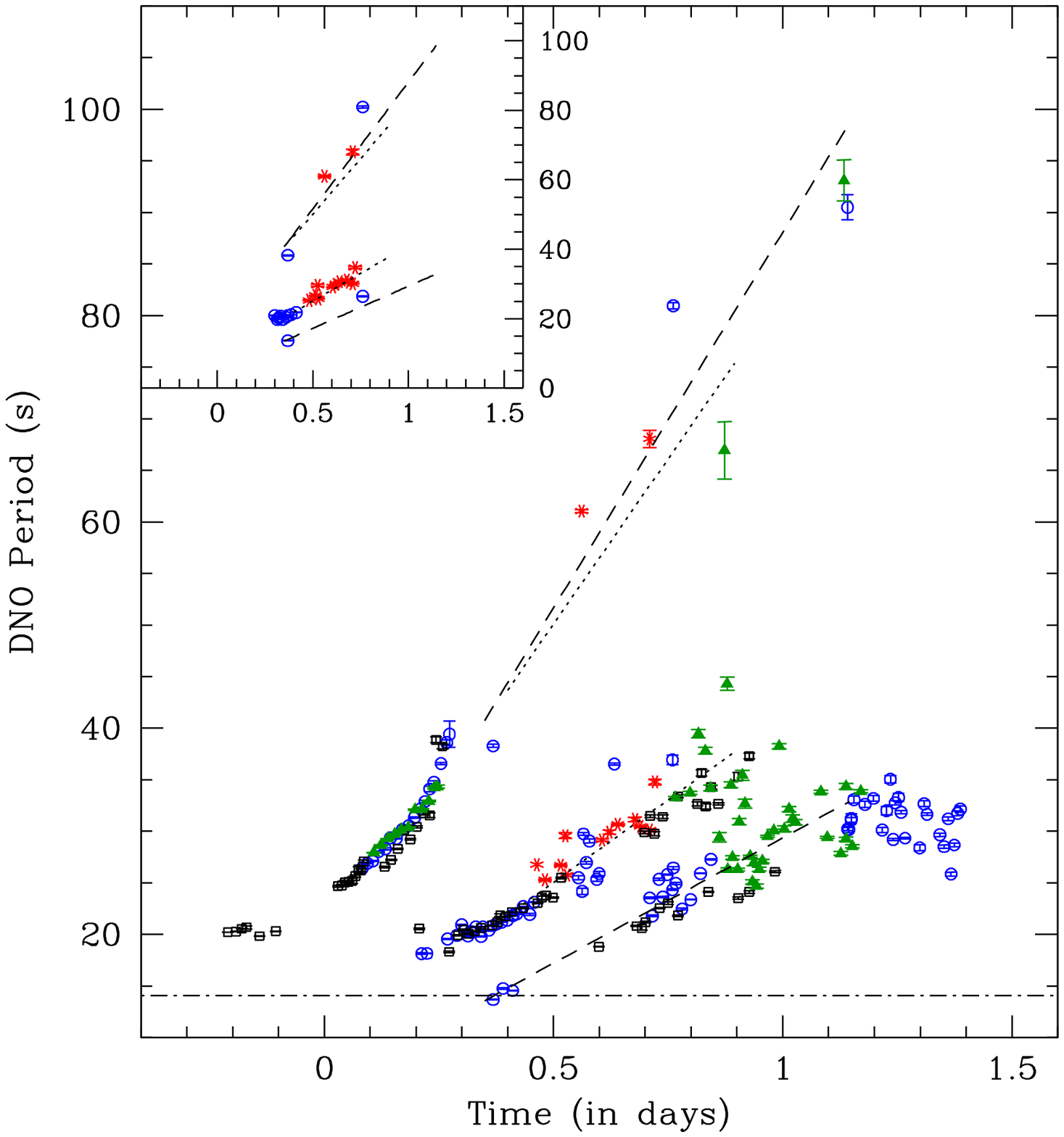,width=8.5cm}}}
  \caption{The evolution of DNO periods at the end of normal and super outbursts
in the dwarf nova VW Hyi. The different symbols indicate the different kind of
outbursts (short: asterisk, normal: open circles, long: open squares and super
outbursts: filled triangles). The dotted and dashed lines show the result of the least 
squares fit to the first and second harmonic, respectively, and are multiplied by a factor
of two and three to show the evolution of the fundamental DNO period. 
The horizontal dotted-dashed line illustrates the minimum
DNO period (14.1 s) observed at maximum brightness. The inset highlights
two observing runs in which the fundamental, first and second harmonic of the DNO period
were occasionally present simultaneously. }
 \label{dno4fig1}
\end{figure}

\begin{figure}
\centerline{\hbox{\psfig{figure=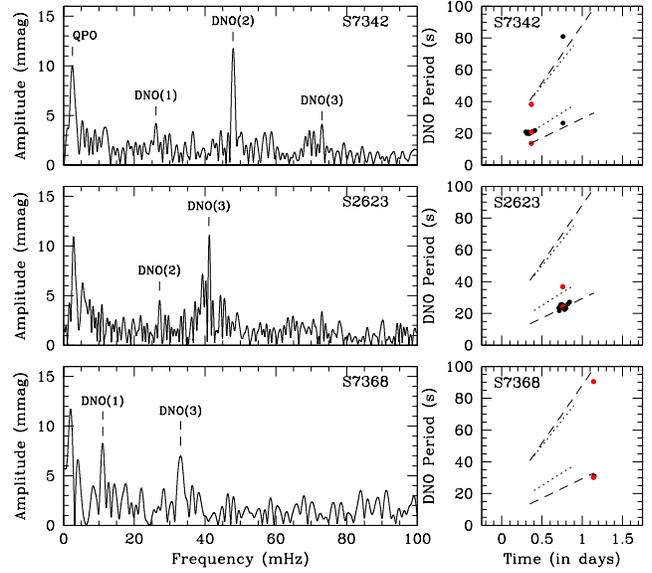,width=8.5cm}}}
  \caption{Examples of parts of runs where there are harmonics simultaneously present. The left panels show
selected Fourier transforms from three observing runs (S7342, S2623, S7368, respectively, from top to bottom); 
prominent peaks are marked and labelled. The right panels
show a small version of Fig.~\ref{dno4fig1} for the three runs displayed here.
}
 \label{dno4fig2}
\end{figure}

     The over all behaviour is displayed in Fig.~\ref{dno4fig1}. The latter uses an 
abscissa on which time $T$ is measured in days before or after a selected 
point near the end of the outburst decay light curve where all the light 
curves coincide; this is defined in Paper I. This enables us to place all 
the DNO evolutions on the same diagram; we estimate that the uncertainty in 
such timing (which arises almost entirely in fitting the observed light curve 
to the standard outburst templates: see Paper I) is $\pm$0.05 d. We have omitted 
most of the slow increase of $P_{DNO}$ before $T$ = 0, which is given in 
Paper I and is typical of the d$P_{DNO}$/d$t$ seen in dwarf novae in general. 
The horizontal dashed line at $P_{DNO}$ = 14.1 s shows the minimum value 
reached at maximum of outburst.

\begin{table*}
 \centering
  \caption{DNOs and QPOs in VW Hyi during decline from normal and super outburst.}
  \begin{tabular}{@{}lrrccclc@{}}
 Run No.   & $<$$T$$>$ & Length         & \multicolumn{3}{c}{DNOs}  &  \multicolumn{2}{c}{QPOs} \\
           & (d)       & (s)            & \multicolumn{3}{c}{(periods in seconds)} & \multicolumn{2}{c}{(period in seconds)}\\
           &           &                & \multicolumn{3}{c}{[amplitude in mmag]} & \multicolumn{2}{c}{[amplitude in mmag]} \\[2pt]
           &           &                & Fundamental                   & First harmonic        & Second harmonic &  &  \\[8pt]
S0018      &  --0.21   &  3105$^*$      & 20.22 $\pm$ 0.02 \hfill [2.2] & \hspace{1.1cm}    -- \hspace{1.1cm}   & \hspace{1.1cm}   -- \hspace{1.1cm}   & &  -- \\
           &  --0.19   &  3455$^*$      & 20.28 $\pm$ 0.02 \hfill [2.2] &     --    &    --    & &  -- \\
           &  --0.18   &  2176$^*$      & 20.55 $\pm$ 0.03 \hfill [2.2] &     --    &    --    & &  -- \\
           &  --0.17   &   448          & 20.70 $\pm$ 0.15 \hfill [4.3] &     --    &    --    & &  -- \\[3pt]
S6133      &  --0.14   &  2418          & 19.84 $\pm$ 0.02 \hfill [1.8] &     --    &    --    & &  -- \\
           &  --0.11   &  3576          & 20.31 $\pm$ 0.02 \hfill [1.4] &     --    &    --    & &  -- \\[3pt]
S6184      &    0.03   &   598          & 24.68 $\pm$ 0.09 \hfill [16.2]&     --    &    --    & &  -- \\
           &    0.04   &   861          & 24.73 $\pm$ 0.10 \hfill  [8.6]&     --    &    --    & &  -- \\
           &    0.05   &   345          & 25.05 $\pm$ 0.22 \hfill [14.2]&     --    &    --    & &  -- \\
           &    0.05   &  1039          & 25.11 $\pm$ 0.05 \hfill [14.7]&     --    &    --    & &  -- \\
           &    0.06   &   343          & 25.22 $\pm$ 0.25 \hfill  [9.7]&     --    &    --    & \vline & \\
           &    0.07   &   861          & 25.66 $\pm$ 0.09 \hfill [10.3]&     --    &    --    & \vline & \\
           &    0.08   &   343          & 26.33 $\pm$ 0.24 \hfill [12.4]&     --    &    --    & \vline & 525 $\pm$ 4 \, [13.6] \\
           &    0.08   &   519          & 26.22 $\pm$ 0.14 \hfill [13.3]&     --    &    --    & \vline & \\
           &    0.09   &   519          & 27.09 $\pm$ 0.12 \hfill [13.6]&     --    &    --    & \vline & \\[3pt]
S6059      &    0.08   &   981          & 26.44 $\pm$ 0.05 \hfill [13.3]&     --    &    --    & \hfill \, \vline & 407 $\pm$ 2 \, [18.3]    \\
           &    0.09   &   976          & 26.90 $\pm$ 0.05 \hfill [14.7]&     --    &    --    & \hfill \vline &     \\
           &    0.11   &   984          & 27.11 $\pm$ 0.05 \hfill [15.2]&     --    &    --    & \vline  \hfill &     \\
           &    0.12   &  1245          & 28.05 $\pm$ 0.05 \hfill [12.5]&     --    &    --    & \vline  \hfill & 442 $\pm$ 3 \, [20.7]    \\
           &    0.13   &   865          & 28.29 $\pm$ 0.11 \hfill [11.9]&     --    &    --    & \vline  \hfill &     \\
           &    0.14   &  1120          & 29.35 $\pm$ 0.13 \hfill  [6.0]&     --    &    --    & \hfill \vline & 430 $\pm$ 6 \, [18.6]    \\
           &    0.16   &  1125          & 29.30 $\pm$ 0.06 \hfill [12.8]&     --    &    --    & &     \\
           &    0.17   &  1032          & 30.18 $\pm$ 0.07 \hfill [10.8]&     --    &    --    & &     \\
           &    0.18   &  1057          & 30.55 $\pm$ 0.06 \hfill [14.6]&     --    &    --    & \vline \hfill & 463 $\pm$ 6 \, [14.0]    \\
           &    0.20   &  1208          & 31.34 $\pm$ 0.07 \hfill [11.8]&     --    &    --    & \vline \hfill &     \\
           &    0.21   &   908          & 32.18 $\pm$ 0.11 \hfill [12.9]&     --    &    --    & &     \\
           &    0.22   &   912          & 32.89 $\pm$ 0.11 \hfill [12.1]&     --    &    --    & &     \\
           &    0.23   &   460          & 34.11 $\pm$ 0.26 \hfill [10.9]&     --    &    --    & \hfill \vline & 578 $\pm$ 10 \, [22.5]    \\
           &    0.24   &  1124          & 34.75 $\pm$ 0.14 \hfill  [9.8]&     --    &    --    & \hfill \vline &     \\
           &    0.25   &   949          & 36.57 $\pm$ 0.12 \hfill [15.0]&     --    &    --    & \vline \hfill &     \\
           &    0.27   &   546          & 38.60 $\pm$ 0.25 \hfill [17.1]&     --    &    --    & \vline \hfill & 285 $\pm$ 4 \, [13.9]    \\
           &    0.27   &   224          & 39.43 $\pm$ 1.26 \hfill [14.0]&     --    &    --    & \vline \hfill &     \\[3pt]
S0127      &    0.11   &  2592$^*$      & 27.94 $\pm$ 0.03 \hfill  [8.8]&     --    &    --    & &  -- \\
           &    0.12   &  2592$^*$      & 28.64 $\pm$ 0.05 \hfill  [5.5]&     --    &    --    & &  -- \\
           &    0.14   &  2592$^*$      & 29.29 $\pm$ 0.04 \hfill  [7.3]&     --    &    --    & &  -- \\
           &    0.15   &  2592$^*$      & 29.72 $\pm$ 0.06 \hfill  [7.8]&     --    &    --    & &  452 $\pm$ 5 \, [19.6]\\
           &    0.17   &  2592$^*$      & 30.15 $\pm$ 0.03 \hfill  [9.2]&     --    &    --    & &  -- \\
           &    0.18   &  2592$^*$      & 30.39 $\pm$ 0.04 \hfill  [7.1]&     --    &    --    & &  -- \\
           &    0.20   &  2592$^*$      & 32.09 $\pm$ 0.06 \hfill  [5.9]&     --    &    --    & &  -- \\
           &    0.21   &  2592$^*$      & 32.14 $\pm$ 0.05 \hfill  [9.4]&     --    &    --    & &  443 $\pm$ 4 \, [22.3] \\
           &    0.23   &  2592$^*$      & 32.94 $\pm$ 0.07 \hfill  [7.1]&     --    &    --    & &  492 $\pm$ 4 \, [27.8] \\
           &    0.24   &  1900$^*$      & 34.24 $\pm$ 0.08 \hfill  [8.9]&     --    &    --    & &  -- \\
           &    0.25   &   604          & 34.33 $\pm$ 0.14 \hfill [13.8]&     --    &    --    & &  -- \\[3pt]
S7621      &    0.13   &  1200          & 26.55 $\pm$ 0.07 \hfill  [9.8]&     --    &    --    & \vline &     \\
           &    0.15   &  1200          & 27.22 $\pm$ 0.05 \hfill [11.0]&     --    &    --    & \vline &     \\
           &    0.16   &  1200          & 28.28 $\pm$ 0.05 \hfill [10.7]&     --    &    --    & \vline & 425 $\pm$ 2 \, [11.2]    \\
           &    0.17   &  1200          & 30.13 $\pm$ 0.05 \hfill  [3.8]&     --    &    --    & \vline &     \\
           &    0.19   &  1200          & 29.22 $\pm$ 0.05 \hfill [12.1]&     --    &    --    & \vline &     \\
           &    0.20   &  1200          & 30.42 $\pm$ 0.05 \hfill [11.1]&     --    &    --    & \vline &     \\
           &    0.22   &  1200          & 31.73 $\pm$ 0.05 \hfill  [7.4]&     --    &    --    & &  --   \\
           &    0.23   &  1200          & 31.55 $\pm$ 0.05 \hfill  [4.9]&     --    &    --    & &  --   \\
           &    0.24   &  1200          & 38.88 $\pm$ 0.05 \hfill  [5.9]&     --    &    --    & &  --   \\
           &    0.26   &  1200          & 38.22 $\pm$ 0.05 \hfill  [8.1]&     --    &    --    & &  --   \\
           &    0.27   &   600          &    --  & 18.32 $\pm$ 0.09 \hfill [4.9]    &    --    & &  --   \\[3pt]
S2915      &    0.21   &   948          &    --  & 20.57 $\pm$ 0.08 \hfill [8.9]    &    --    & &  -- \\[3pt]
S1307      &   0.21    &   600          &         --     &  18.15 $\pm$ 0.08 \hfill [5.0] &    --  & & 198  $\pm$ 2 \ \, [4.4]\\
           &   0.22    &  1032          &         --     &  18.16 $\pm$ 0.04 \hfill [3.9] &    --   & &  -- \\[3pt]
\end{tabular}
{\footnotesize 
\newline 
$^*$ This section includes a 50\% overlap with the subsequent section.\\
}
\label{dno4tab2}
\end{table*}
\addtocounter{table}{-1}

\begin{table*}
 \centering
  \caption{Continued: DNOs and QPOs in VW Hyi during decline from normal and super outburst.}
  \begin{tabular}{@{}lrrccclc@{}}
 Run No.   & $<$$T$$>$ & Length         & \multicolumn{3}{c}{DNOs}  &  \multicolumn{2}{c}{QPOs} \\
           & (d)       & (s)            & \multicolumn{3}{c}{(periods in seconds)} & \multicolumn{2}{c}{(period in seconds)}\\
           &           &                & \multicolumn{3}{c}{[amplitude in mmag]} & \multicolumn{2}{c}{[amplitude in mmag]} \\[2pt]
           &           &                & Fundamental                   & First harmonic        & Second harmonic & &   \\[8pt]
S7311      &   0.25    &  2000          &         --                      &            --                  &                           --   & & 399 $\pm$ 4 \ \, [9.9] \\
           &   0.27    &  1036          &         --                      &  19.57 $\pm$ 0.04 \hfill [6.2] &                           --   & &  --  \\
           &   0.29    &  2333          &         --                      &  19.93 $\pm$ 0.01 \hfill [6.1] &                           --   & &  -- \\
           &   0.31    &   692          &         --                      &  20.38 $\pm$ 0.11 \hfill [3.8] &                           --   & &  -- \\
           &   0.32    &  1512          &         --                      &  20.27 $\pm$ 0.02 \hfill [8.6] &                           --   & & 298 $\pm$ 4 \, [10.1] \\
           &   0.35    &  3241          &         --                      &  20.75 $\pm$ 0.01 \hfill [7.7] &                           --   & \vline \hfill & \\
           &   0.38    &  1900          &         --                      &  21.05 $\pm$ 0.01 \hfill [14.0]&                           --   & \vline \hfill & \\
           &   0.39    &   432          &         --                      &          --                    &  14.77 $\pm$ 0.08 \hfill [8.6] & \vline \hfill & \\
           &   0.40    &  1124          &         --                      &  21.37 $\pm$ 0.05 \hfill [10.4]&                           --   & \vline \hfill & 351 $\pm$ 6 \ \, [8.7] \\
           &   0.41    &  1036          &         --                      &  21.85 $\pm$ 0.04 \hfill [11.5]&  14.56 $\pm$ 0.04 \hfill [5.5] & \vline \hfill & \\
           &   0.42    &   504          &         --                      &  22.01 $\pm$ 0.07 \hfill [22.2]&                           --   & \vline \hfill & 657 $\pm$ 3 \ \, [6.6] \\
           &   0.43    &  1384          &         --                      &  22.71 $\pm$ 0.02 \hfill [15.2]&                           --   & \vline \hfill & \\
           &   0.45    &   432          &         --                      &  21.96 $\pm$ 0.09 \hfill [16.6]&                           --   & \vline \hfill & \\
           &   0.46    &  1480          &         --                      &  23.13 $\pm$ 0.02 \hfill [12.0]&                           --   & \vline \hfill & \\[3pt]
S7342      &   0.30    &  1576          &         --                      &  20.94 $\pm$ 0.04 \hfill [3.6] &                           --   & &  254 $\pm$ 2 \ \, [9.4] \\
           &   0.31    &   780          &         --                      &  19.85 $\pm$ 0.07 \hfill [7.4] &                           --   & &  -- \\
           &   0.32    &   690          &         --                      &  20.24 $\pm$ 0.09 \hfill [6.0] &                           --   & &  -- \\
           &   0.33    &   780          &         --                      &  20.73 $\pm$ 0.04 \hfill [9.8] &                           --   & &  -- \\
           &   0.34    &  1230          &         --                      &  19.82 $\pm$ 0.04 \hfill [5.2] &                           --   & &  262 $\pm$ 5 \ \, [9.5] \\
           &   0.36    &   565          &         --                      &  20.40 $\pm$ 0.07 \hfill [11.3]&                           --   & &  -- \\
           &   0.37    &   990          &  38.28 $\pm$ 0.15 \hfill [4.3]  &  20.86 $\pm$ 0.03 \hfill [11.7]&  13.69 $\pm$ 0.04 \hfill [4.1] & &  -- \\
           &   0.39    &  2250          &         --                      &  21.20 $\pm$ 0.02 \hfill [7.7] &                           --   & &  -- \\
           &   0.41    &  1195          &         --                      &  21.82 $\pm$ 0.04 \hfill [11.0]&                           --   & &  -- \\[3pt]
S6316      &   0.29    &   951          &         --                      &  19.92 $\pm$ 0.04 \hfill [6.4] &                           --   & &  -- \\
           &   0.30    &   995          &         --                      &  20.47 $\pm$ 0.05 \hfill [7.0] &                           --   & &  -- \\
           &   0.31    &   950          &         --                      &  20.06 $\pm$ 0.04 \hfill [7.0] &                           --   & &  -- \\
           &   0.33    &  1470          &         --                      &  20.33 $\pm$ 0.03 \hfill [6.8] &                           --   & &  -- \\
           &   0.34    &   950          &         --                      &  20.62 $\pm$ 0.09 \hfill [3.9] &                           --   & &  -- \\
           &   0.37    &  1555          &         --                      &  20.87 $\pm$ 0.03 \hfill [7.7] &                           --   & &  -- \\
           &   0.38    &   430          &         --                      &  21.20 $\pm$ 0.09 \hfill [12.0]&                           --   & &  -- \\
           &   0.38    &   650          &         --                      &  21.86 $\pm$ 0.13 \hfill [6.3] &                           --   & &  -- \\
           &   0.40    &  1424          &         --                      &  21.74 $\pm$ 0.03 \hfill [9.1] &                           --   & &  226 $\pm$ 2 \, [11.6] \\
           &   0.41    &   950          &         --                      &  22.18 $\pm$ 0.04 \hfill [9.4] &                           --   & &  -- \\
           &   0.43    &  3283          &         --                      &  22.56 $\pm$ 0.03 \hfill [2.1] &                           --   & &  -- \\
           &   0.47    &   950          &         --                      &  23.04 $\pm$ 0.05 \hfill [8.6] &                           --   & &  -- \\
           &   0.47    &   560          &         --                      &  23.56 $\pm$ 0.17 \hfill [6.1] &                           --   & &  -- \\
           &   0.48    &  1081          &         --                      &  23.77 $\pm$ 0.05 \hfill [10.0]&                           --   & &  -- \\
           &   0.50    &  1555          &         --                      &  23.56 $\pm$ 0.06 \hfill [4.0] &                           --   & &  231 $\pm$ 2 \ \, [9.2] \\
           &   0.52    &  1450          &         --                      &  25.49 $\pm$ 0.07 \hfill [5.4] &                           --   & &  -- \\[3pt]
S6138      &   0.49    &  1516          &         --                      &  25.31 $\pm$ 0.06 \hfill [6.8] &                           --   & \vline \, \vline & 285 $\pm$ 4 \, [28.3] \\
           &   0.52    &  1474          &         --                      &  26.70 $\pm$ 0.09 \hfill [7.9] &                           --   & \vline \hfill &     \\
           &   0.53    &   437          &         --                      &  29.58 $\pm$ 0.29 \hfill [11.4]&                           --   & \vline \hfill & 615 $\pm$ 2 \, [26.0]  \\
           &   0.54    &  1209          &         --                      &  25.77 $\pm$ 0.09 \hfill [8.9] &                           --   & \vline \hfill &     \\
           &   0.57    &  2977          &  61.05 $\pm$ 0.17 \hfill [10.7] &            --                  &                           --   & &  -- \\
           &   0.61    &  1690          &         --                      &  29.16 $\pm$ 0.08 \hfill [7.5] &                           --   & &  -- \\
           &   0.63    &  1042          &         --                      &  29.92 $\pm$ 0.18 \hfill [7.8] &                           --   & &  -- \\
           &   0.65    &  2295          &         --                      &  30.66 $\pm$ 0.07 \hfill [7.1] &                           --   & & 350 $\pm$ 3 \, [30.6] \\
           &   0.68    &   610          &         --                      &  31.16 $\pm$ 0.21 \hfill [11.6]&                           --   & &  -- \\
           &   0.70    &  1517          &         --                      &  30.49 $\pm$ 0.06 \hfill [9.8] &                           --   & \vline \hfill &   \\
           &   0.72    &  1128          &  68.04 $\pm$ 0.83 \hfill [8.8]  &  30.13 $\pm$ 0.12 \hfill [12.3]&                           --   & \vline \hfill &  384 $\pm$ 2 \, [31.1] \\
           &   0.73    &   955          &         --                      &  34.78 $\pm$ 0.28 \hfill [6.9] &                           --   & \vline \hfill &  768 (subharm.)\\[3pt]
S7384      &   0.46    &  2108          &         --                      &  26.76 $\pm$ 0.05 \hfill [6.4] &                           --   & &  -- \\[3pt]
S3416      &   0.56    &   478          &         --                      &  25.50 $\pm$ 0.21 \hfill [8.4] &                           --   & \vline \hfill &   \\
           &   0.56    &   346          &         --                      &  24.19 $\pm$ 0.31 \hfill [12.1]&                           --   & \vline \hfill &  333 $\pm$ 2 \, [28.4] \\
           &   0.57    &   431          &         --                      &  26.96 $\pm$ 0.12 \hfill [17.6]&                           --   & \vline \hfill &   \\
           &   0.58    &   736          &         --                      &  29.07 $\pm$ 0.19 \hfill [9.2] &                           --   & \vline \hfill &   \\
           &   0.59    &   492          &         --                      &  29.75 $\pm$ 0.12 \hfill [16.6]&                           --   & &  -- \\
           &   0.59    &   492          &         --                      &  25.36 $\pm$ 0.22 \hfill [10.2]&                           --   & &  -- \\
           &   0.60    &   384          &         --                      &  25.91 $\pm$ 0.21 \hfill [16.4]&                           --   & &  -- \\
\end{tabular}
\end{table*}
\addtocounter{table}{-1}

\begin{table*}
 \centering
  \caption{Continued: DNOs and QPOs in VW Hyi during decline from normal and super outburst.}
  \begin{tabular}{@{}lrrccclc@{}}
 Run No.   & $<$$T$$>$ & Length         & \multicolumn{3}{c}{DNOs}  &  \multicolumn{2}{c}{QPOs} \\
           & (d)       & (s)            & \multicolumn{3}{c}{(periods in seconds)} & \multicolumn{2}{c}{(period in seconds)}\\
           &           &                & \multicolumn{3}{c}{[amplitude in mmag]} & \multicolumn{2}{c}{[amplitude in mmag]} \\[2pt]
           &           &                & Fundamental                   & First harmonic        & Second harmonic &  &  \\[8pt]
S6528      &   0.60    &  1904          &         --                      &              --                &  18.81 $\pm$ 0.03 \hfill [6.0] & &  -- \\
           &   0.68    &  1728          &         --                      &              --                &  20.81 $\pm$ 0.05 \hfill [3.4] & &  -- \\
           &   0.69    &   476          &         --                      &              --                &  20.60 $\pm$ 0.19 \hfill [5.4] & &  -- \\
           &   0.70    &   905          &         --                      &              --                &  21.20 $\pm$ 0.08 \hfill [5.0] & &  -- \\
           &   0.73    &  2028          &         --                      &              --                &  22.56 $\pm$ 0.07 \hfill [3.1] & &  -- \\
           &   0.75    &   932          &         --                      &              --                &  23.04 $\pm$ 0.12 \hfill [4.0] & &  -- \\
           &   0.84    &   932          &         --                      &              --                &  24.13 $\pm$ 0.07 \hfill [5.7] & &  -- \\
           &   0.90    &   864          &         --                      & 35.32 $\pm$ 0.39 \hfill [4.2]  &  23.53 $\pm$ 0.12 \hfill [5.9] & &  -- \\
           &   0.93    &  1249          &         --                      & 37.29 $\pm$ 0.28 \hfill [3.8]  &  24.12 $\pm$ 0.12 \hfill [3.6] & &  -- \\
           &   0.98    &  1573          &         --                      &              --                &  26.09 $\pm$ 0.10 \hfill [3.0] & &  -- \\[3pt]
S5248      &   0.63    &  3455          &         --                      & 36.53 $\pm$ 0.09 \hfill [17.8] &              --                & &  -- \\[3pt]
S0019      &   0.70    &  1242          &         --                      & 29.90 $\pm$ 0.14 \hfill [9.9]  &              --                & \vline \hfill &     \\
           &   0.71    &  1035          &         --                      & 31.48 $\pm$ 0.12 \hfill [16.4] &              --                & \vline \hfill& 514 $\pm$ 2 \, [35.5] \\
           &   0.72    &   520          &         --                      & 29.76 $\pm$ 0.19 \hfill [21.0] &              --                & \vline \hfill&     \\
           &   0.74    &   995          &         --                      & 31.42 $\pm$ 0.18 \hfill [10.2] &              --                & \vline \hfill&     \\
           &   0.77    &  1340          &         --                      & 33.39 $\pm$ 0.13 \hfill [11.8] &  21.83 $\pm$ 0.09 \hfill [7.8] & &  -- \\
           &   0.81    &   865          &         --                      & 32.68 $\pm$ 0.20 \hfill [11.0] &              --                & &  -- \\
           &   0.82    &   690          &         --                      & 35.68 $\pm$ 0.31 \hfill [12.2] &              --                & &  -- \\
           &   0.83    &   865          &         --                      & 32.40 $\pm$ 0.25 \hfill [11.4] &              --                & &  -- \\
           &   0.84    &  1125          &         --                      & 34.27 $\pm$ 0.16 \hfill [13.3] &              --                & &  -- \\
           &   0.86    &  1725          &         --                      & 32.65 $\pm$ 0.10 \hfill [12.2] &              --                & &  -- \\[3pt]
S2623      &   0.71    &   928          &         --                      &             --                 &  23.55 $\pm$ 0.07 \hfill [12.9]& \vline \hfill&     \\
           &   0.72    &   730          &         --                      &             --                 &  21.81 $\pm$ 0.07 \hfill [18.5]& \vline \hfill&     \\
           &   0.73    &  1015          &         --                      &             --                 &  25.36 $\pm$ 0.09 \hfill  [9.3]& \vline \hfill&     \\
           &   0.74    &   450          &         --                      &             --                 &  23.63 $\pm$ 0.07 \hfill [18.8]& \vline \hfill&  948 $\pm$ 5 \, [18.4] \\
           &   0.75    &  1230          &         --                      &             --                 &  25.80 $\pm$ 0.06 \hfill [10.6]& \vline \hfill&     \\
           &   0.76    &   760          &         --                      & 36.93 $\pm$ 0.45 \hfill [5.0]  &  24.35 $\pm$ 0.08 \hfill [11.0]& \vline \hfill&     \\
           &   0.77    &   585          &         --                      &             --                 &  24.95 $\pm$ 0.11 \hfill [12.0]& \vline \hfill&     \\
           &   0.78    &  1660          &         --                      &             --                 &  22.48 $\pm$ 0.10 \hfill  [6.1]&  & --    \\
           &   0.80    &  1640          &         --                      &             --                 &  23.38 $\pm$ 0.03 \hfill [15.1]& \vline \hfill&  \\
           &   0.82    &  2075          &         --                      &             --                 &  25.92 $\pm$ 0.03 \hfill  [8.8]& \vline \hfill& 256 $\pm$ 1 \, [12.6] \\
           &   0.84    &  1895          &         --                      &             --                 &  27.27 $\pm$ 0.05 \hfill  [9.2]& \vline \hfill &  \\[3pt]
S7343      &   0.76    &   750          &  80.97 $\pm$ 0.30 \hfill [8.5]  &             --                 &  26.45 $\pm$ 0.12 \hfill  [8.9]& &  -- \\[3pt]
S0484      &   0.77    &  1396          &         --                      & 33.25 $\pm$ 0.15 \hfill [5.5]  &                     --         & \vline \hfill&     \\
           &   0.80    &  1816          &         --                      & 33.72 $\pm$ 0.14 \hfill [6.2]  &                     --         & \vline \hfill&     \\
           &   0.82    &  1380          &         --                      & 39.48 $\pm$ 0.41 \hfill [3.9]  &                     --         & \vline \, \vline&     \\
           &   0.83    &  1288          &         --                      & 37.81 $\pm$ 0.34 \hfill [3.6]  &                     --         & \vline \, \vline& 457 $\pm$ 3 \, [31.8] \\
           &   0.84    &   608          &         --                      & 34.21 $\pm$ 0.28 \hfill [7.9]  &                     --         & \vline \, \vline&     \\
           &   0.86    &   520          &         --                      & 29.43 $\pm$ 0.45 \hfill [8.7]  &                     --         & \vline \hfill&     \\
           &   0.87    &   196          &  66.93 $\pm$ 2.78 \hfill [12.4] &          --                    &                     --         & \vline & 1328 $\pm$ 2 \, [44.1] \\
           &   0.88    &   780          &         --                      & 44.31 $\pm$ 0.63 \hfill [7.6]  &                     --         & \vline \hfill&     \\
           &   0.89    &   732          &         --                      & 34.51 $\pm$ 0.27 \hfill [6.3]  &                     --         & \vline \, \vline&     \\
           &   0.91    &   692          &         --                      & 30.92 $\pm$ 0.30 \hfill [6.3]  &                     --         & \vline \, \vline& 530 $\pm$ 4 \, [36.1] \\
           &   0.91    &   388          &         --                      & 35.44 $\pm$ 0.49 \hfill [8.6]  &                     --         & \vline \, \vline&     \\
           &   0.92    &   604          &         --                      & 32.69 $\pm$ 0.41 \hfill [6.3]  &                     --         & \vline \, \vline&     \\[3pt]
S7222      &   0.88    &   790          &         --                      &            --                  &  26.34 $\pm$ 0.14 \hfill [8.8] & &  -- \\
           &   0.89    &   584          &         --                      &            --                  &  27.47 $\pm$ 0.18 \hfill [10.3]& &  -- \\
           &   0.90    &  1380          &         --                      &            --                  &  26.30 $\pm$ 0.12 \hfill  [5.3]& &  -- \\
           &   0.93    &   948          &         --                      &            --                  &  27.59 $\pm$ 0.11 \hfill  [7.5]& &  -- \\
           &   0.93    &   288          &         --                      &            --                  &  25.08 $\pm$ 0.20 \hfill  [7.5]& &  -- \\
           &   0.94    &   432          &         --                      &            --                  &  26.99 $\pm$ 0.34 \hfill  [6.2]& &  -- \\
           &   0.94    &   432          &         --                      &            --                  &  24.68 $\pm$ 0.20 \hfill  [7.7]& &  -- \\
           &   0.95    &   608          &         --                      &            --                  &  26.34 $\pm$ 0.20 \hfill  [7.1]& &  -- \\
           &   0.96    &   688          &         --                      &            --                  &  27.12 $\pm$ 0.18 \hfill  [5.2]& &  -- \\
           &   0.97    &  1184          &         --                      &            --                  &  29.51 $\pm$ 0.11 \hfill  [7.9]& \vline \hfill &     \\
           &   0.98    &  1152          &         --                      &            --                  &  30.08 $\pm$ 0.12 \hfill  [7.6]& \vline \hfill &   271 $\pm$ 4 \, [11.9] \\
           &   0.99    &   944          &         --                      & 38.24 $\pm$ 0.25 \hfill [7.3]  &               --               & \vline \hfill &     \\
           &   1.00    &   908          &         --                      &            --                  &  30.25 $\pm$ 0.18 \hfill [10.3]& &  -- \\
           &   1.01    &   992          &         --                      &            --                  &  32.15 $\pm$ 0.24 \hfill  [6.7]& &  -- \\
\end{tabular}
\end{table*}
\addtocounter{table}{-1}

\begin{table*}
 \centering
  \caption{Continued: DNOs and QPOs in VW Hyi during decline from normal and super outburst.}
  \begin{tabular}{@{}lrrccclc@{}}
 Run No.   & $<$$T$$>$ & Length         & \multicolumn{3}{c}{DNOs}  &  \multicolumn{2}{c}{QPOs} \\
           & (d)       & (s)            & \multicolumn{3}{c}{(periods in seconds)} & \multicolumn{2}{c}{(period in seconds)}\\
           &           &                & \multicolumn{3}{c}{[amplitude in mmag]} & \multicolumn{2}{c}{[amplitude in mmag]} \\[2pt]
           &           &                & Fundamental                   & First harmonic        & Second harmonic &  &  \\[8pt]
S7222      &   1.02    &   348          &         --                      & \hspace{1cm} -- \hspace{1cm}   &  31.25 $\pm$ 0.37 \hfill [10.9]& &  -- \\
           &   1.03    &   656          &         --                      &              --                &  30.88 $\pm$ 0.24 \hfill [6.2] & &  -- \\
           &   1.08    &  1036          &         --                      &              --                &  33.84 $\pm$ 0.14 \hfill [9.6] & &  -- \\
           &   1.10    &  1467          &         --                      &              --                &  29.41 $\pm$ 0.10 \hfill [6.8] & & 306 $\pm$ 2 \, [23.4]\\
           &   1.13    &   864          &         --                      &              --                &  27.86 $\pm$ 0.19 \hfill [8.1] & &  -- \\
           &   1.14    &  1081          &         --                      &              --                &  29.25 $\pm$ 0.19 \hfill [8.9] & &  -- \\
           &           &                &         --                      &              --                &  34.36 $\pm$ 0.28 \hfill [8.2] & &  -- \\
           &   1.15    &  1380          &         --                      &              --                &  28.56 $\pm$ 0.15 \hfill [6.2] & &  -- \\
           &   1.17    &  1728          &         --                      &              --                &  33.88 $\pm$ 0.14 \hfill [5.8] & &  -- \\[3pt]
S0129      &   1.13    &  3885          &  93.15 $\pm$ 0.29 \hfill [16.1] &              --                &              --                & &  -- \\[3pt]
S7368      &   1.14    &   800          &  90.49 $\pm$ 1.23 \hfill  [8.2] &              --                &  30.23 $\pm$ 0.16 \hfill [7.0] & \vline &  321 $\pm$ 3 \, [24.5] \\
           &   1.15    &   573          &         --                      &              --                &  31.11 $\pm$ 0.23 \hfill [9.2] & \vline &   \\[3pt]
S7301      &   1.14    &   520          &         --                      &              --                &  30.19 $\pm$ 0.31 \hfill [7.6] & &  -- \\
           &   1.15    &   475          &         --                      &              --                &  31.26 $\pm$ 0.42 \hfill [7.6] & &  -- \\
           &   1.16    &   430          &         --                      &              --                &  33.08 $\pm$ 0.34 \hfill [8.1] & &  -- \\
           &   1.18    &   775          &         --                      &              --                &  32.62 $\pm$ 0.31 \hfill [8.5] & &  -- \\
           &   1.20    &  1165          &         --                      &              --                &  33.19 $\pm$ 0.29 \hfill [6.2] & &  -- \\
           &   1.22    &   910          &         --                      &              --                &  30.13 $\pm$ 0.24 \hfill [8.4] & &  -- \\
           &   1.23    &   735          &         --                      &              --                &  31.99 $\pm$ 0.40 \hfill [5.0] & &  -- \\
           &   1.24    &   690          &         --                      &              --                &  35.04 $\pm$ 0.36 \hfill [6.3] & &  -- \\
           &   1.25    &  1210          &         --                      &              --                &  32.78 $\pm$ 0.20 \hfill [5.9] & &  -- \\
           &   1.26    &   935          &         --                      &              --                &  31.85 $\pm$ 0.16 \hfill [7.7] & &  -- \\
           &   1.31    &   635          &         --                      &              --                &  32.68 $\pm$ 0.28 \hfill [7.5] & &  -- \\
           &   1.41    &  7236          &         --                      &              --                &               --               & & 581 $\pm$ 2 \, [19.0]    \\[3pt]
S1322      &   1.24    &   880          &         --                      &              --                &  29.21 $\pm$ 0.11 \hfill [13.8]& &  -- \\
           &   1.25    &   348          &         --                      &              --                &  33.29 $\pm$ 0.37 \hfill [11.0]& &  -- \\
           &   1.27    &  2072          &         --                      &              --                &  29.34 $\pm$ 0.08 \hfill [6.8] & &  -- \\
           &   1.30    &   520          &         --                      &              --                &  28.41 $\pm$ 0.29 \hfill [8.9] & &  -- \\
           &   1.31    &  1036          &         --                      &              --                &  31.66 $\pm$ 0.13 \hfill [8.0] & &  -- \\
           &   1.34    &   656          &         --                      &              --                &  29.66 $\pm$ 0.17 \hfill [18.2]& \vline &     \\
           &   1.35    &   944          &         --                      &              --                &  28.52 $\pm$ 0.16 \hfill [11.2]& \vline &     \\
           &   1.36    &   648          &         --                      &              --                &  31.22 $\pm$ 0.19 \hfill [13.4]& \vline &  424 $\pm$ 1 \, [29.7] \\
           &   1.37    &   404          &         --                      &              --                &  25.85 $\pm$ 0.19 \hfill [20.0]& \vline &     \\
           &   1.37    &   804          &         --                      &              --                &  28.68 $\pm$ 0.16 \hfill [8.3] & \vline &     \\
           &   1.38    &   484          &         --                      &              --                &  31.75 $\pm$ 0.23 \hfill [14.2]& &  -- \\
           &   1.39    &   468          &         --                      &              --                &  32.15 $\pm$ 0.29 \hfill [12.4]& &  -- \\[3pt]
\end{tabular}
\end{table*}

    Discussing first the results from the FT analyses, which are biased towards the stronger 
signals but show the gross trends, we see that from $T$ = 0 to $T$ = 0.20 d there is a monotonic 
single-valued increase in $P_{DNO}$ from 25 s to 33 s, as already seen in Paper I but 
now with additional data. At $T$ = 0.20 d the first runs appear in which a 1st harmonic 
is sometimes present, but only at $T$ = 0.25 d does the 1st harmonic dominate over 
the fundamental. The apparent turn-up in the trend of the fundamental in its last few points 
may be due to the unrecognised presence of synodic rather than direct DNOs, but we see no evidence 
for any double DNO in the observations at this stage (in the rare cases when we observed a double 
DNO at other times we have plotted only the direct period in Fig.~\ref{dno4fig1}).

     From $T$ = 0.25 to 0.50 the 1st harmonic increases in period with little scatter, but from 
$T$ = 0.36 it is occasionally accompanied by a 2nd harmonic, which to our knowledge is the first 
such observation of frequency tripling in a CV.

       From $T$ = 0.5 to 1.0 d the evolution of the DNOs is characterised by increasing domination 
of the 2nd harmonic over the 1st, until at $T$ = 1.0 only the 2nd harmonic is present and 
appears to stabilize at a period $\sim$ 30 s, though with considerable scatter. For a day or so 
after $T$ = 1.4 d, by which time VW Hyi is fully at quiescence, our light curves show no 
DNOs at all. On rare occasions through the $T$ = 0.25 -- 1.1 d range there are guest 
appearances of the fundamental.

     In most of the runs after $T$ = 0.3 d either the 1st or the 2nd harmonic is present, 
and there is alternation between them within the run, but in runs S6138 and S7342 there are 
parts where fundamental and both harmonics are simultaneously present in the FTs; these 
are shown in the inset in Fig.~\ref{dno4fig1}. Our FTs show that the 1:2:3 ratios 
of periods are satisfied within errors of measurement when the components are present together. 
In Fig.~\ref{dno4fig2} we illustrate FTs that contain combinations of the fundamental, 
2nd and 3rd harmonics present simultaneously (because one or other component usually 
dominates, the presence of the weaker components is usually only made convincing by 
inspecting FTs of the immediately preceding or subsequent sections of the runs, 
where they in turn may dominate).

   In Fig.~\ref{dno4fig1} the dotted and dashed lines are least squares fits 
to the 1st and 2nd harmonics (with all parameters free, but omitting the 2nd harmonic 
points with $T$ $>$ 1.15 d) and appear again in the upper part of the diagram multiplied 
by two and three respectively to show the approximate evolution of the implied fundamental.

\begin{figure}
\centerline{\hbox{\psfig{figure=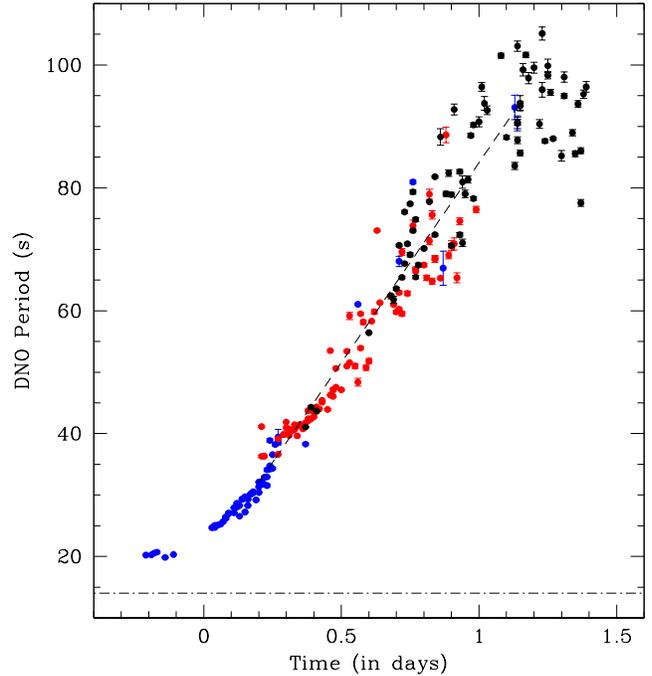,width=8.5cm}}}
  \caption{Time evolution of the observed or implied fundamental DNO period.}
 \label{dno4fig3}
\end{figure}

   Fig.~\ref{dno4fig3} shows a modified form of Fig.~\ref{dno4fig1}, where we have now 
transformed the 1st and 2nd harmonics into the implied fundamental to show its 
evolution more clearly. This shows that the fundamental systematically increases in period 
to $\sim$ 105 s before the DNOs disappear completely. When passing through 85 -- 95 s any 
appearance of the fundamental in the FT may be confused with the lpDNOs in VW Hyi that 
occur in that period range. However, the lpDNOs observed so far (in VW Hyi and other CVs) 
are all pure sinusoids and often long-lasting, so most of them can be recognised for what 
they are and have been omitted from Fig.~\ref{dno4fig1}. In VW Hyi it therefore appears 
that at times the fundamental period of the normal DNOs can briefly reach values 
greater than the period of the lpDNOs, but the tendency for the 2nd harmonic to 
stabilize near $\sim$ 30 s at the end of the whole DNO evolution is a strong indication 
that $P_{DNO} \rightarrow P_{lpDNO}$ at the end of outburst.

   The dashed line in Fig.~\ref{dno4fig3} is the least squares fit for $T$ = 0.2 to 1.15 d 
and has the equation

\begin{equation}
       P_{DNO} (s) = 65.06 \ (\pm \, 1.49) \  T\,(d) \  + \  19.0 \ (\pm \, 1.0).
\label{dno4eq1}
\end{equation}

The formal errors on the parameters derived from the least squares fit are given in Eq.~\ref{dno4eq1}
in brackets.

\subsubsection{Some details of the DNO behaviour}

    In order to obtain greater sensitivity to the fast changing periods we have calculated 
amplitude and O--C values, fitting sinusoids by least squares to short and usually 
overlapping sections of the light curves. These are amplitude/phase diagrams, which we 
will call A/$\phi$ plots. Here we will show a few of the more significant ones.

\begin{figure}
\centerline{\hbox{\psfig{figure=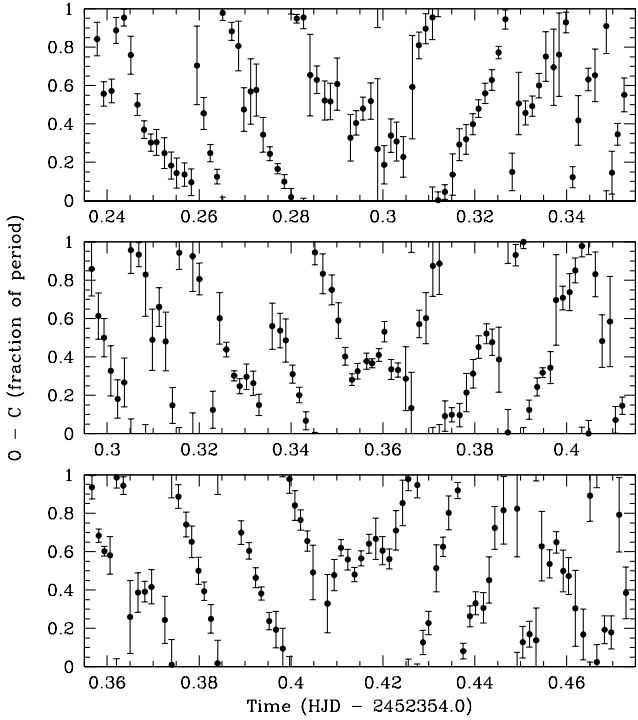,width=8.5cm}}}
  \caption{Phase diagram for the first harmonic component of DNOs in run S6316. The three panels
are for the first half (relative to $P_{DNO} = 20.62$ s), the centre half (22.18 s), and the final half 
(23.04 s) of the run. The phases `wrap around' if they exceed one period.}
 \label{dno4fig4}
\end{figure}

\begin{figure}
\centerline{\hbox{\psfig{figure=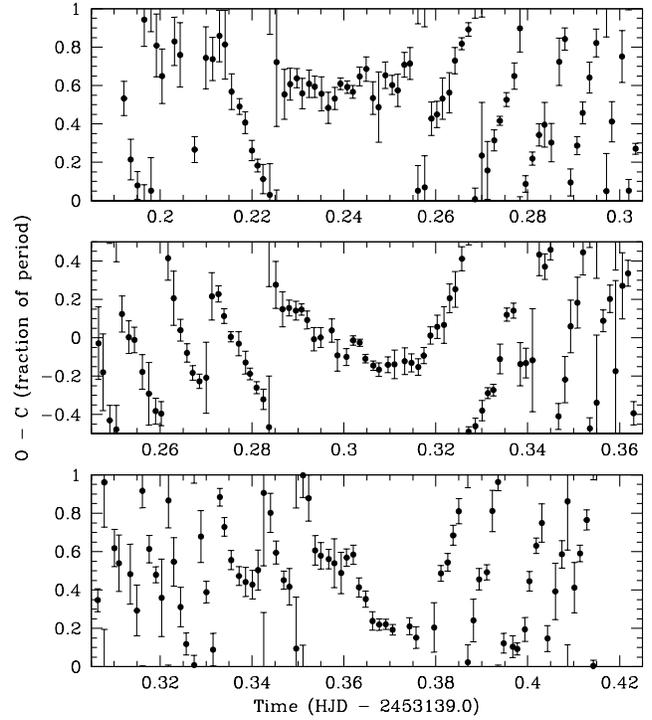,width=8.5cm}}}
  \caption{As for Fig.~\ref{dno4fig4}, but for run S7311. The phase diagrams are plotted relative to 
a DNO period of 19.93 s (top), 20.75 s (middle) and 22.01 s (bottom), respectively.}
 \label{dno4fig5}
\end{figure}

   The FT of run S6316 contains only 1st harmonic DNOs and S7311 is predominantly 
1st harmonic. In order to compare their behaviour with that of the fundamental (which 
would have been strongly present only an hour or so before the start of these runs -- as 
seen in S6059 in Table~\ref{dno4tab2}) we give in Figs.~\ref{dno4fig4} and \ref{dno4fig5} phase diagrams 
structured in the same manner as the `oak panel diagram' of S6059 shown in figure 10 
of Paper I.
These show that the general behaviour of fundamental and 1st harmonic DNOs 
are essentially identical: short-lived increases and decreases of period or phase 
superposed on the steady increase in mean period. S7311 is, however, relatively 
more coherent than S6316, with smaller deviations from the secular increase or period.

\begin{figure}
\centerline{\hbox{\psfig{figure=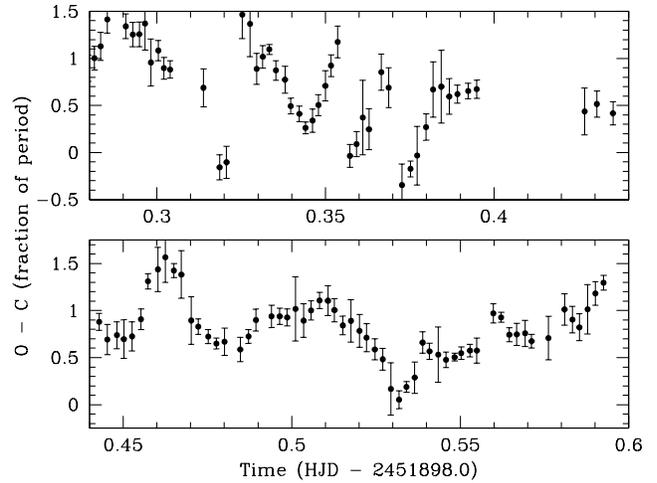,width=8.5cm}}}
  \caption{Phase variations of the first harmonic in run S6138.}
 \label{dno4fig7}
\end{figure}

   Run S6138 is predominantly 1st harmonic and is the first to show noticeable scatter 
in period. To examine this at greater resolution we give a phase diagram in Fig.~\ref{dno4fig7}, 
where the run has been divided into two equal parts and the phases are measured 
relative to periods 25.77 s and 30.50 s respectively. In the upper panel there is the 
quasi-parabolic variation with phase changing by several cycles, which is the signature of 
a steadily increasing period, but in the lower panel there is variation (with a 
range $\sim$ one cycle) around a constant phase, showing no systematic increase in 
period over the 3.7 h span of time. There is no noticeable difference in the gross light 
curve behaviour between the two halves of the run (the second half of which is shown in 
figure 2 of Paper II), but we note that optical flux is largely determined by accretion 
in the outer parts of the disc, whereas the evident temporary relative stability of $P_{DNO}$ 
and its implied $\dot{M}$ is a property of the inner disc.

\begin{figure}
\centerline{\hbox{\psfig{figure=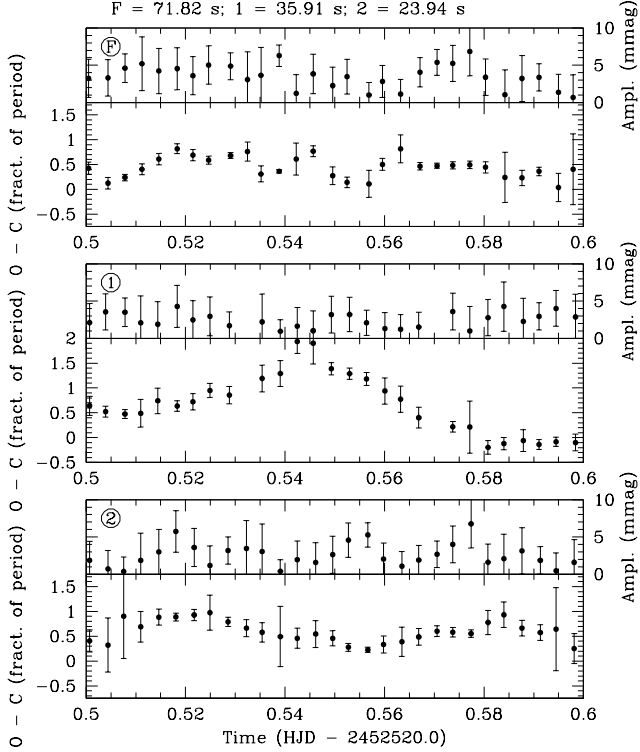,width=8.5cm}}}
  \caption{Phase and amplitude variations of the fundamental (top), first harmonic (middle)
and second harmonic (bottom) in run S6528.}
 \label{dno4fig8}
\end{figure}

   Towards the end of run S6528, which again is largely 2nd harmonic, the FTs show 
occasional detectable presence of the 1st harmonic (see Table~\ref{dno4tab2}). We use this 
run to demonstrate the importance of not relying on FTs alone. The bottom panel of 
Fig.~\ref{dno4fig8} shows the A/$\phi$ diagram for the 2nd harmonic, relative to a test 
sinusoid with period 23.94 s, which we have smoothed and reduced the uncertainties 
by using longer sections of light curve. Despite the low amplitude of the signal, 
the smoothly varying phase measurements show that the 2nd harmonic is almost 
always present. The middle panel shows the A/$\phi$ diagram for the 
1st harmonic, using a test sinusoid of period ${{3}\over{2}} \times 23.94 = 35.91$ s, where 
it can be seen that the 1st harmonic is also almost always present, but its phase variations 
are not obviously positively or negatively correlated with those of the 2nd harmonic; 
this means, of course, that they do not maintain an exact 2:3 period ratio, but nor do 
they drift apart in any systematic way. Finally, in the top panel of Fig.~\ref{dno4fig8} we 
show the result of searching for the presence of the fundamental, using $3 \times 23.94 = 71.82$ s 
as the comparison period. This reveals that, despite there being no clear peak in the FT 
near 71.82 s (which is a region of increased noise at the lower frequency end of CVs 
in general), there is a low amplitude fundamental almost continuously present, with 
well-defined phase that appears to follow the variation of the 2nd harmonic.

    We have investigated sections of other runs where the periods stay almost constant 
for considerable lengths of time -- these give similar impressions to what we have found 
in run S6528, but with less certainty because of smaller amplitudes.

\begin{figure}
\centerline{\hbox{\psfig{figure=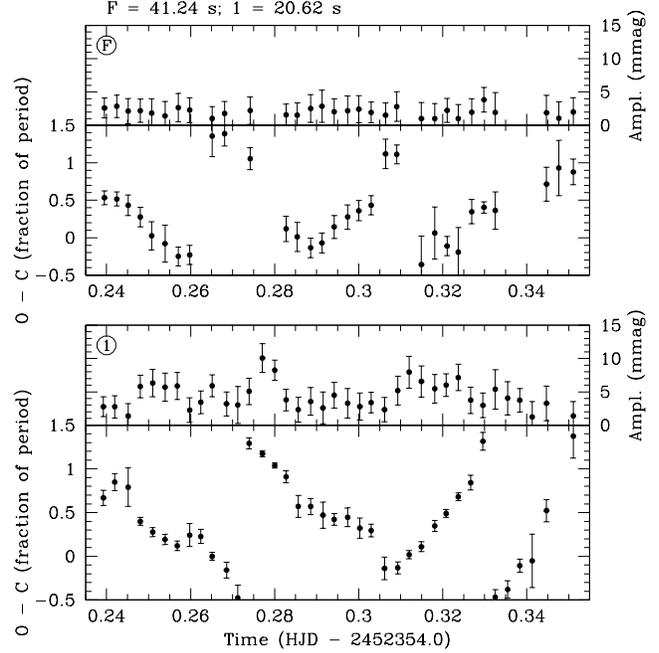,width=8.5cm}}}
  \caption{Phase and amplitude variations of the implied fundamental (top), and observed first harmonic (bottom)
in run S6316.}
 \label{dno4fig9}
\end{figure}

  There remains to show what DNO components can be discerned when the period is 
changing rapidly. We analyse two runs as exemplars. First we look again at run S6316, for 
which Fig.~\ref{dno4fig4} gives the oak panel diagram of the 1st harmonic. In Fig.~\ref{dno4fig9} 
the first half of the run is shown, giving both A and $\phi$ in the lower panel, relative to 
a sinusoid of period 20.62 s but using longer sections of light curve to provide a smoother version 
of the top panel of Fig.~\ref{dno4fig4}. The upper panel of Fig.~\ref{dno4fig9} shows the result 
of least squares fits in the neighbourhood of the implied fundamental (i.e., using a 
test period of 41.24 s). It is seen that the fundamental is continually present, albeit 
of very low amplitude, and that it has a `parabolic' phase variation similar to that 
of the 1st harmonic (note that because the period is double, the total phase variation 
of the fundamental is half that of the 1st harmonic).

\begin{figure}
\centerline{\hbox{\psfig{figure=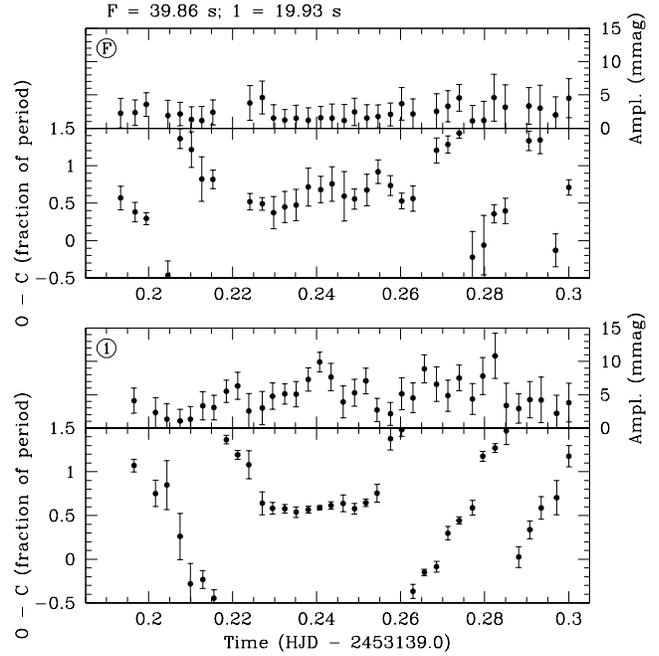,width=8.5cm}}}
  \caption{Phase and amplitude variations of the implied fundamental (top), and observed first harmonic (bottom)
in run S7311.}
 \label{dno4fig10}
\end{figure}

   Lastly we look at run S7311, which is the first run after the fundamental has 
disappeared and the 1st harmonic is strong (see Fig.~\ref{dno4fig1}). In Fig.~\ref{dno4fig10} 
the 1st harmonic shows a parabolic phase change and there is a low amplitude fundamental 
with the same behaviour. On the other hand, the 2nd harmonic does not appear to be 
present -- the amplitudes are very small and the phases spread over a full cycle 
and do not show the parabolic variation.

   In summary, for small variation, the fundamental and 2nd harmonic components follow each other in phase 
variation, but the 1st harmonic has an independent variation. For the more general steady increase in period
the fundamental and harmonics follow each other as expected.

\subsection{The Behaviour of the QPOs}

\begin{figure}
\centerline{\hbox{\psfig{figure=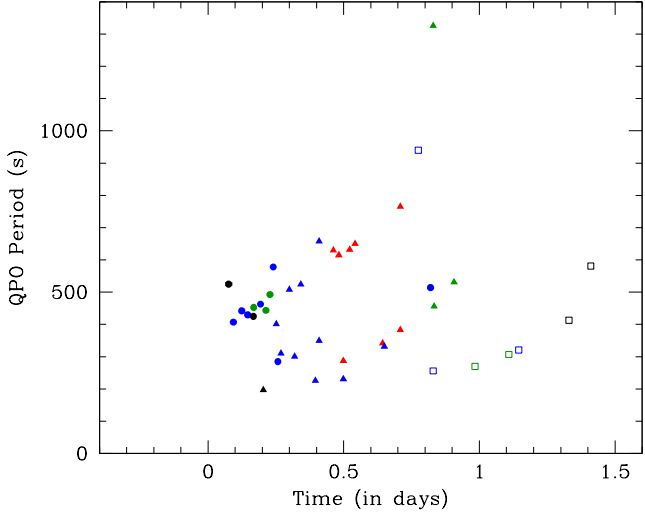,width=8.5cm}}}
  \caption{The evolution of QPO periods in the late decline of outbursts in VW Hyi. The filled circles indicate runs with
the fundamental DNO period predominantly present, filled triangles represent runs where the first harmonic of the DNO period 
is dominant, and the open squares indicate runs where the second harmonic of the DNO is present.}
 \label{dno4fig11}
\end{figure}

    We have analysed the QPO signals in the same way as for the DNOs. The QPOs are initially
identified in the Fourier transform of subsections of data for each observing run; these subsections have typical data lengths of $\sim$30 to $\sim$45 minutes,
corresponding to between $\sim$5 and $\sim$10 cycles of a QPO modulation. Given the incoherent nature of the QPOs
and the presence of flickering on similar time-scales, 
QPOs are difficult to identify using standard techniques. Therefore, we have been fairly conservative in identifying the 
QPOs -- only those that are then clearly seen in the light curves for at least several cycles 
are accepted. There are probably other short-lived QPOs, made difficult to identify 
because of stochastic flickering in the light curve.
In Paper I we reported 
that the $\sim$ 600 s QPOs in the February 2000 outburst of VW Hyi halved in period near 
the end of our run (see also Table~\ref{dno4tab2}). This was a first hint that we should expect frequency doubling in the QPOs, 
as observed in the DNOs. This has proved to be the case, and also extends to tripling of 
frequencies -- our results are shown in Fig.~\ref{dno4fig11} and the periods are listed 
in Table~\ref{dno4tab2}.

\begin{figure}
\centerline{\hbox{\psfig{figure=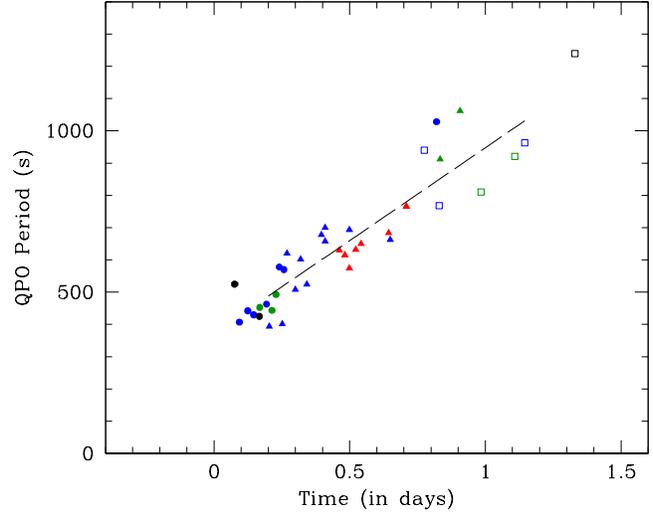,width=8.5cm}}}
  \caption{The evolution of the measured and implied fundamental QPO periods in VW Hyi. The symbols are as in Fig.~\ref{dno4fig11}
and the dashed line is a least squares fit to the data (see Eq.~\ref{dno4eq2}).}
 \label{dno4fig12}
\end{figure}

    From Fig.~\ref{dno4fig11} we see that the first appearance of the first harmonic of the 
QPOs is at $T$ = 0.2 d, which is the same place at which the DNO 1st harmonic begins. 
The second harmonic starts at $T \sim 0.8$ d, which is where the DNO second harmonics begin
to become very prominent. Again we can produce a modified plot, given in Fig.~\ref{dno4fig12}, in 
which we have shown the implied fundamental QPO periods. The solid line in Fig.~\ref{dno4fig12} is 
a least squares fit to these points and has the equation, for the range 0.2 $<$ $T(d)$ $<$ 1.15,

\begin{equation}
       P_{QPO} (s) = 575 \ (\pm \, 54) \ T\, (d) \  + \  372 \ (\pm \, 33),
\label{dno4eq2}
\end{equation}

where the formal errors on the parameters derived from the least squares fit are given in brackets 
in the above equation (Eq.~\ref{dno4eq2}).

   Equations~\ref{dno4eq1} and \ref{dno4eq2} show that the ratio $R$ is not constant 
during the factor of three increase in fundamental period seen in VW Hyi. In fact, $R$ changes 
from 15.2 at $T$ = 0.2 d to 11.0 at $T$ = 1.15 d, which is typical of the range of values seen 
among other CVs, e.g., Mauche (2002) found $R$ = 10.4 and 11.4 for two runs on SS Cyg and in 
Paper III we found $R$ = 15 for TU Men and V893 Sco. Values of $R$ outside of this 
range may arise from unwittingly combining DNO and QPO fundamentals with harmonics.

\section{Interpretation}

\subsection{Introduction}

    In order to prepare for our suggested interpretation of the complicated 
behaviour of VW Hyi we recall some of the structural components that have been 
used to explain the rapid brightness modulations in CVs.

\begin{enumerate}
\item{DNOs seen at optical wavelengths have their origin in the reprocessing of harder radiation, 
emitted anisotropically from the accretion zone(s) near the primary, striking the primary or 
sweeping around the concave surface of the accretion disc. These generate a modulation 
at the rotation period of the accretion zone, $P_{rot}$. It is possible also to generate 
a signal at ${{1}\over{2}} P_{rot}$: although the simplest model of a tilted concave disc 
generates a sinusoidal signal, with the far side and near side disc contributions 
being 180$^{\circ}$ out of phase and therefore giving an amplitude which is the 
difference of the two signals (Petterson 1980), more generally the two components may not be 
so completely sinusoidal. For example, if there were strong forward or backward scattering 
then the signal would have narrow peaks when the beam was pointing directly forward or away 
from us. Furthermore, the signals from the far side of the disc and from the primary are 
180$^{\circ}$ out of phase, which can generate two peaks per rotation. There may also 
be situations where two accretion zones are visible (to the disc or to us) which also 
produce what is in effect the 1st harmonic of the rotation period.}
\item{If the rotating beam intercepts a structure such as the secondary, or the thickened 
disc in the vicinity of the stream impact on the disc, which is revolving with the orbital 
period, then a modulation is generated at the `synodic' period $P_{syn}$, obtained from 
$P_{syn}^{-1}$ = $P_{rot}^{-1}$ -- $P_{orb}^{-1}$, which we can more compactly write in 
frequencies as $\omega_{syn} = \omega_{rot} - \Omega_{orb}$.  This process is commonly 
present in intermediate polars and generates a single orbital sideband to the $\omega_{rot}$ signal.}
\item{If the reprocessing site is periodically modulated in cross section, this introduces 
an amplitude modulation into the reprocessed signal. As an example, suppose that the reprocessing 
site is most favourably placed just once per orbital period, this then generates sidebands 
to the $\omega_{syn}$ signal at $\omega_{syn} + \Omega_{orb}$ and $\omega_{syn} - \Omega_{orb}$, 
i.e. signals at $\omega_{rot}$ and $\omega_{rot} - 2 \Omega_{orb}$, as well as the 
original  $\omega_{rot} - \Omega_{orb}$ (see Warner (1986) for this and more complicated 
examples). The sideband signals, which would be of equal amplitude in the simplest case, are 
often unequal because of the presence of components generated by other mechanisms, which may be 
in or out of phase with them (and may be of variable amplitude).}
\item{If $\dot{M}$ onto the primary is itself periodically modulated then sidebands 
to the $\omega_{rot}$ signal or any of the reprocessed signals may be generated. This 
is the `beat frequency' model introduced first for CVs (Warner 1983), shown later not to 
work for the standard DNOs (Warner 1987), but independently invented and successfully 
applied to some types of QPOs in X-Ray Binaries (Alpar \& Shaham 1985; Lamb et al.~1985), 
and used to interpret T Tauri light curves (Smith, Bonnell \& Lewis 1995).}
\end{enumerate}

\subsection{The DNO Frequency Doubling}

   There are at least two ways in which frequency doubling of DNOs might arise. Given 
the general structure of the LIMA model, one possibility is a transition from single-pole to 
two-pole accretion. This might be due simply to a change in the field-threading region 
as $\dot{M}$ changes, resulting in feeding of two accretion zones instead of one. 
A transition to quadrupole field from dipole field configuration may occur when the inner 
edge of the accretion disc is pushed so close to the white dwarf surface that the less 
compressible higher order multipole field components determine the position of the inner 
edge of the disc (Lamb 1988); but in reverse this would reduce the number of accretion 
zones as $\dot{M}$ decreases and is thus unlikely to be related to frequency doubling 
in VW Hyi late in outburst when $\dot{M}$ is decreasing.

     It is not easy to make quantitative predictions about the changes in magnetic 
field structure that would lead to the above effect, but another possibility is that 
frequency doubling arises through a geometrical rearrangement, which allows us (or 
the reprocessing regions of the accretion disc) to view two poles instead of one. The 
smoothness of the transition to a dominant doubled DNO frequency -- in the sense that 
there is no concomitant apparent change in behaviour (coherence, jump to longer periods, 
amplitude) -- suggests that no radical restructuring has occurred in the threading region 
or in the accretion zones. We have therefore considered models in which either the 
upper accretion zone becomes visible as the inner edge of the accretion disc retreats 
outwards with decreasing $\dot{M}$, or the second accretion zone becomes visible. 
Such a transition is well documented in the case of the X-Ray light curve of 
the intermediate polar XY Ari (Hellier, Mukai \& Beardmore 1997).

   In Paper I we noted that the shortest observed $P_{DNO}$ in VW Hyi is 14.1 s 
(seen only very rarely, but in both optical and X-Rays at the maxima of different 
outbursts). Interpreted as the period of Keplerian rotation at the surface of the 
primary (i.e. the minimum period that the innermost parts of the disc or the 
equatorial belt can have before magnetic channeling of gas ceases as the 
magnetosphere of the primary is crushed by the mass flow) this implies a primary 
mass $M(1) = 0.70$ M$_{\odot}$ and radius $R(1) = 7.79 \times 10^8$ cm from 
the Nauenberg (1972) mass-radius relationship for white dwarfs. From the large 
amplitude orbital humps but absence of eclipses in the light curve of VW Hyi it 
has long been assumed that the inclination is $\sim 60^{\circ}$. However, on 
2 January 2004, towards the end of a superoutburst of VW Hyi, we observed what 
appear to be grazing eclipses of the bright spot, made possible by the increased 
size of the accretion disc as it expanded in response to the outward flow of angular 
momentum from the gas falling inwards. A similar effect has been seen in TY PsA 
(Haefner, Schoembs \& Vogt 1979; Warner, O'Donoghue \& Wargau 1989). We therefore 
use the slightly larger inclination of $64^{\circ}$.

     In the LIMA model the accretion zones are within the equatorial band and 
will be long arcs, probably one somewhat above the equator and one below, on 
opposite sides of the primary. When the inner edge of the disc is close to the primary, 
the primary itself obscures the inner parts of the disc on the far side. As the 
inner radius $r_{in}$ of the disc increases with decreasing $\dot{M}$ in the final 
phase of an outburst, these inner regions, which contain the upper accretion 
curtain, come into view beyond the primary (and at the same time part of the 
accretion curtain feeding the second pole may become visible beyond the inner 
edge of the disc on the near side). This happens at $r_{in} \sim R(1) \sec i$. 
The Keplerian period of the inner edge of the disc is then found from  

\begin{equation}
        P_{k}^{2} = 4 \pi^{2} \, R^{3}(1) \, \sec^{3}i \, /\, G M(1),
\end{equation}

which produces 46.2 s $<$ $P_k$ $<$ 51.4 s for the range $63^{\circ} < i < 65^{\circ}$. 
We suggest, therefore, that for $P_{DNO} < P_{k}$ we view only the `upper' accretion zone 
and curtain when it is on our side of the primary, but when $P_{DNO} > P_{k}$ we see that 
accretion curtain again when it is on the far side of the primary (where it may appear 
even brighter: Hellier et al.~1987); and there may also be a contribution from the 
lower accretion zone, both giving a double humped luminosity profile for each rotation 
of the equatorial belt. Any inequality of apparent luminosity of the two components 
will add a fundamental period to the 1st harmonic.

     Frequency doubling should therefore occur near $P_{DNO} \sim 48$ s, as is indeed observed.

    We note also that frequency doubling of a synodic DNO can occur simply by a QPO 
traveling wave itself transforming to a double-humped profile, i.e. through excitation of the 1st harmonic.

\subsection{The Appearance of a Second Harmonic}

   Frequency tripling is not a common physical phenomenon -- in particular we think it very 
unlikely that in VW Hyi it could be the result of accretion onto three magnetic regions. We 
therefore consider a model in which the 2nd harmonic of the DNO fundamental does not represent 
a physical frequency in the VW Hyi system, but rather is, just as observed, a Fourier component 
of the light curve. The fact that the 2nd harmonic does not appear until the 1st harmonic is 
already present is in agreement with this hypothesis, though this cannot be a strict 
prerequisite because the 2nd harmonic on its own appears in many of the later light curves. 
When this happens it implies that there are other sources of fundamental and 1st harmonic 
signals that can be out of phase with the amplitude modulated components at those frequencies.

\subsubsection{Beat Frequency from the Wall}

     Consider the state of VW Hyi nearing quiescence after outburst. The disc has a 
very low $\dot{M}$ and will be thin in vertical extent. When the fundamental $P_{DNO} \sim 60$ s 
the inner disc has been emptied by the primary's rotating magnetosphere out to $\sim 1 \times 10^9$ cm 
above the surface of the primary. The structure subtending the largest angular size that 
the rotating DNO beam can illuminate is the thickened part of the inner disc that generates 
the QPO modulation (we note here that QPOs are a common feature in these late stages of a 
VW Hyi outburst). The QPO traveling wave will also effectively block illumination by the 
DNO beam of a large area of the disc, but the disc will always be accessible to the beam 
during the half of the beam cycle that passes over the lower half of the wave profile. Depending 
on the relative strengths of these two components, the rapid oscillations seen in these late 
phases of outburst could at times be predominantly synodic DNOs rather than direct ones.

    The postulated QPO wave is a region of disc thickening near the inner edge of the 
accretion disc. The strongest field lines, which will be the dominant collectors of gas 
from the disc, sweep past this thicker region\footnote{The region of disc thickening will 
not necessarily be where the mass transfer is highest -- the thickest part of the wave 
might (because of mass continuity in the Keplerian flow) be a region of relative rarefaction 
and cause a lower rate of mass transfer. But in any case there is a periodic modulation 
of mass transfer as the field lines sweep around the inner edge of the disc.} with a 
frequency $\omega_{rot} - \omega_{qpo}$, where $\omega_{qpo}$ is the frequency of 
the traveling wave (typically about one fifteenth of $\omega_{rot}$, unless frequency 
doubling has occurred in one of them). We therefore may expect the two accretion zones 
on the primary each to be modulated in luminosity at a frequency $\omega_{rot} - \omega_{qpo}$.

   The result is that the 1st harmonic, generated either as a result of seeing one luminous 
zone twice per rotation, or two accretion zones separated in longitude by roughly $180^{\circ}$, 
is modulated in luminosity. For synodic DNOs the 1st harmonic has frequency 
2 $(\omega_{rot} - \omega_{qpo}$) and the $\dot{M}$ modulation frequency is 
$\omega_{rot} - \omega_{qpo}$. This will generate Fourier components that include 
sidebands at the sum and difference frequencies (see, e.g., Warner 1986), i.e. 
at $\omega_{rot} - \omega_{qpo}$, 2 $(\omega_{rot} - \omega_{qpo}$) and 
3 $(\omega_{rot} - \omega_{qpo}$). As already mentioned above, there may be other 
mechanisms contributing to the brightness of the fundamental and/or the 1st harmonic 
components, which will alter the amplitudes of those components from the simple 1:2 
ratio of amplitude modulation, but the 2nd harmonic is generated in our proposed 
model solely as a rotational sideband.

\subsubsection{Beat Frequency from the Disc}

    Another possibility is that the traveling wall does not intercept much of 
the beam radiation, but the accretion rate is still modulated at 
frequency $\omega_{rot} - \omega_{qpo}$. The optical DNOs are then generated in 
the normal way off the disc. No longer are there precise harmonic ratios -- we 
would see the frequency set $\omega_{rot} + \omega_{qpo}$, 2 $\omega_{rot}$ and 
3 $\omega_{rot} - \omega_{qpo}$, which differ slightly from 1:2:3 ratios. We 
have not detected any convincing examples of such behaviour in our VW Hyi observations.

\subsubsection{Combinations of the two Beat Frequency Models}

The differences between the models can in principle be tested by looking at the period 
ratios for FTs where the fundamental and its apparent harmonics coexist. There is also 
the possibility of both processes acting simultaneously -- with part of the beamed radiation 
coming from the disc and part from the wall. This would, from comparison of the 
frequencies generated in the two models, produce double DNOs in all components, with 
a frequency splitting of 2 $\omega_{qpo}$. This is different from the non-modulated 
mass transfer case, where the splitting is $\omega_{qpo}$ in the fundamental (Paper II).

    In addition to these possibilities it may happen that there is 
no $\dot{M}$ variation, so the fundamental and its first harmonic can be seen both 
at direct and synodic frequencies, i.e. double DNOs at $\omega_{rot}$, $\omega_{rot} - \omega_{qpo}$ 
and/or at 2 $\omega_{rot}$ and 2 $(\omega_{rot} - \omega_{qpo}$).

    Such simultaneous or alternating presence of the various options could 
account for part or all of the vertical scatter in Figs.~\ref{dno4fig1} and \ref{dno4fig3}. 
It is difficult to test for all of these effects simultaneously, but we may find some 
of these processes at work individually. We give two examples in detail, selected 
from regions in Fig.~\ref{dno4fig1} where there are widely different $P_{DNO}$ in 
contiguous subsections -- in effect, double DNOs with the components separated in time.

\begin{figure}
\centerline{\hbox{\psfig{figure=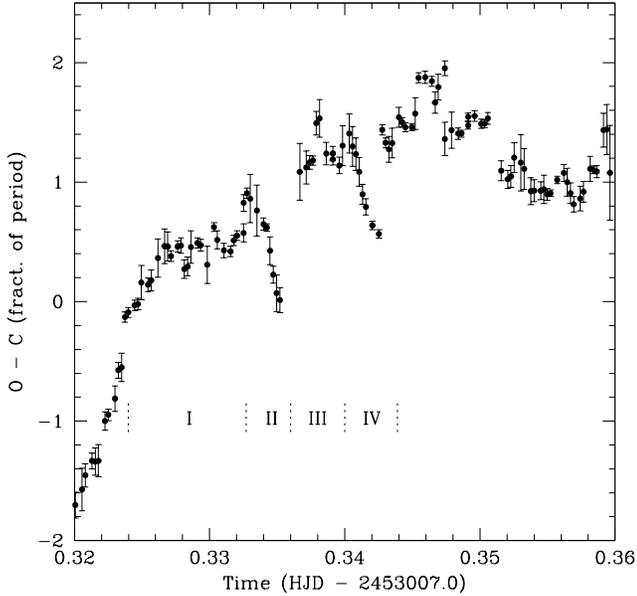,width=8.5cm}}}
  \caption{Phase variation in a section of run S7222.}
 \label{dno4fig13}
\end{figure}

   Fig.~\ref{dno4fig13} shows the A/$\phi$ diagram for a section of Run S7222, computed 
with the test period 27.0 s. The marked subsections I to IV have $P_{DNO}$ values of 
27.59, 25.08, 26.99 and 24.68 s respectively. The beat period between the periods in 
subsections I and II is 285 s, and that for subsections III and IV is 288 s. 
There is no QPO directly detectable in this section of the light curve (although the 
light curve is not particularly noisy at this time), but in the part almost 
immediately following, namely during 0.3750 -- 0.3925 d, there is a strong QPO with a 
period of 271 $\pm$ 4 s. The effect in the A/$\phi$ diagram is, therefore, one that would 
be expected of alternating direct and synodic DNOs, where the DNO beam is 
reprocessed from a QPO traveling wave, but not of sufficient amplitude to produce QPOs 
as seen from our direction. After subsection IV in Fig.~\ref{dno4fig13} there is no 
clear interpretation of the behaviour.

\begin{figure}
\centerline{\hbox{\psfig{figure=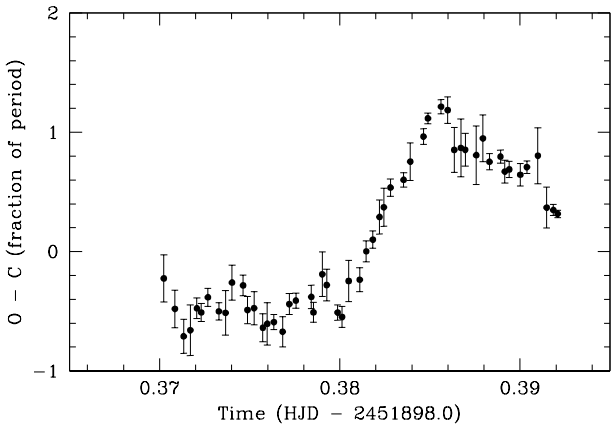,width=8.5cm}}}
  \caption{Phase variation in a section of run S6138.}
 \label{dno4fig14}
\end{figure}

    Fig.~\ref{dno4fig14} is a section of run S6138, with the A/$\phi$ diagram 
computed for the test period 26.70 s. Here we see the reverse behaviour of what happens 
in Run S7222: the sudden change from 26.70 s to 29.58 s, seen as the abrupt change in 
slope at 0.380 d, would imply a change from direct DNO to synodic DNO from a traveling wave 
with a period of 274 s. There is a QPO in the range 0.349 -- 0.363 d with a period of 
285 $\pm$ 3 s, which is not visible to us during the 0.370 -- 0.400 d time frame.

\subsection{QPO Harmonics}

  In Paper II we suggested that the QPO signal arises from a traveling wave in the inner disc, 
obscuring and/or reflecting light from the central source. This is supported by the 
appearance of synodic DNOs, explained as reprocessing by the traveling wave. The existence of 
first and second harmonics to the QPOs does not have a geometrical explanation of the sort we 
invoke for the DNO harmonics; it appears more likely that the excitation mechanism for the 
traveling wave changes as the inner edge of the disc recedes to regions of weaker field strengths.
 
  This in turn could suggest an alternative model for the behaviour of the DNOs. If the excitation 
mechanism of the traveling wave manages to excite the first or second harmonic of the 
wave it will produce a double or triple humped QPO profile. The rotating DNO beam will then 
generate harmonic synodic DNOs at twice or three times the fundamental synodic DNO frequency. 
The appearance of the 1st and 2nd harmonic DNOs at about the same times as the corresponding 
QPO harmonics is in harmony with this model.

    This model has the advantage of automatically preserving the $P_{QPO}/P_{DNO} \sim 15$ 
relationship, but less in its favour is that we sometimes see 1st and 2nd harmonic DNOs 
simultaneously, which requires similar excitation of both harmonics of the traveling wave. 
The rapid apparent jumps from direct to synodic DNOs, seen in the previous Section, are 
also not a simple consequence of this model. But the existence of the QPO harmonics does 
allow, perhaps even demand, some contribution to the DNO harmonics, and we add this 
to the m\'elange already outlined above.

\subsection{Discussion}

   As pointed out in Paper III, the measured $v \sin i$ for the white dwarf in VW Hyi, 
which in combination with a mass and therefore radius of the white dwarf primary 
provides a rotation period, suggests that the lpDNOs are caused by two-pole accretion 
onto the surface of the primary, i.e. the rotation period of the primary is $\sim$ 170 s. 
We hypothesized that variations of $P_{lpDNO}$ can occur because of feeding gas onto 
different magnetic field lines that connect to the surface of the differentially 
rotating primary. Our conclusion here that the ordinary DNOs seen in the late phases 
of outburst involve two accretion zones, and that the angular velocity of the region 
containing those zones is slowed to approximately the same value as that of the 
underlying white dwarf is in accord with this interpretation of the lpDNOs.

   At first sight it might be thought that there is incompatibility between 
our interpretation of the increase of $P_{DNO}$ in the final stage of outburst, as a 
rapid deceleration of the equatorial belt, and the HST observations of a belt that 
is still rotating near Keplerian velocity days or weeks after outburst (Section 1). 
However, in the LIMA model the magnetic field (other than the weak field of the primary) 
is generated by shearing in the accretion flow onto and within the belt, so 
there is a selection effect at work -- the accretion spots that create the DNOs 
are regions of maximum field strength, which are not necessarily regions of high 
angular velocity but rather are in regions of highest shear. In the generation 
of the field and the efficiency of magnetically channeled accretion there is a 
positive feedback mechanism -- accretion torque increases shear which in turn 
enhances the field and ensures continual and increasing capture of accreting gas.

    The phase of strong deceleration in VW Hyi implies strong shear and it is 
therefore not surprising that DNOs of greatest amplitude appear at that time. 
The disappearance of DNOs at $T \sim 1.4$ d may therefore be the result of most 
of the belt having become nearly uniform in angular velocity, near that of the 
underlying star, leaving only a narrow belt of rapid rotation and a relatively 
small shear zone that loses angular momentum much more slowly (through friction with 
the stellar surface and through now weak magnetic coupling to the disc). This could 
also be the prime reason why magnetically controlled accretion, with its DNOs signature, 
is rare in quiescent dwarf novae.

    VW Hyi remains unique as the only dwarf nova showing DNOs throughout the whole 
of its outburst. It would be useful to find another dwarf nova for which similar 
phenomena can be studied and compared with VW Hyi. G\"ansicke, Beuermann \& Thomas (1997) 
have pointed out that many of the general properties of EK TrA strongly resemble 
those of VW Hyi, though the latter has an $\dot{M}$ about five times larger. 
EK TrA is not listed in the CVs that are known to have DNOs (Warner 2004). Recognizing 
that it would be of great value to have a second dwarf nova that shows DNOs right down 
to the beginning of quiescence we observed EK TrA at the end of its May 2004 outburst, 
at $m_V \sim 14.9$, but no DNOs were present.

\section{QPOs IN X-RAY BINARIES}

     Much of the phenomenology of rapid oscillations in CVs described above 
resembles the behaviour of QPOs in X-Ray binaries (XRBs), so we will give a brief 
overview of the latter. Two distinct frequency domains exist among those XRBs that 
possess QPOs (which are often only intermittent): high frequency QPOs that are 
at $\sim$ hundreds of Hz and low frequency QPOs that are at $\sim$ tens of Hz. 
The hfQPOs usually are double, with the separation varying far less than the 
frequencies themselves (as $\dot{M}$ changes). These appear to be the equivalent 
of double DNOs in CVs, but the frequency separation is close to the rotation 
frequency of the compact object, rather than to a more slowly moving traveling 
wave as in the CVs.

    During X-Ray bursts on neutron star XRBs a third QPO frequency domain can appear, 
lying midway between the hfQPOs and lfQPOs. This contains oscillations variable in frequency 
that have been attributed to expanding gas (heated by thermonuclear runaway) on the surface 
of the neutron star, which slips relative to the surface and therefore does not represent 
precisely the rotation period of the star (van der Klis 2000, 2005). These are similar 
in behaviour to the lpDNOs that we see in CVs.

   In XRBs, both for neutron star and black hole systems, the hfQPOs are at $\sim$ 15 
times the frequency of the lfQPOs (Belloni et al.~2002) and maintain this value as 
variations in $\dot{M}$ change the frequencies in the neutron star binaries 
(in black hole binaries\footnote{Note that most of the discrepant
points in the correlation for black hole binaries have been removed -- see
Klu\'zniak et al.~(2005).} the frequencies are almost invariant in each system 
(McClintock \& Remillard 2005) but this may be due to the activity taking place in
the vicinity of the innermost stable circular orbit (van der Klis 2005)). 
Identifying hfQPOs as the equivalent 
to DNOs in CVs, and lfQPOs as equivalent to the CV QPOs, we see the similarity in 
behaviour of XRBs and CVs, including the response to changing $\dot{M}$ in VW Hyi, 
as was first pointed in Paper II and generalized in Paper III.

  A further aspect of hfQPOs in XRBs is that some systems show harmonic structures (Remillard et al.~2002). 
Specific examples are XTE\,J1550-564 which has strong modulations near 276 and 184 Hz and 
some power at 92 Hz, which are in the ratios 3:2:1, GRO\,J1655-40 which has 
450 and 300 Hz oscillations (Strohmayer 2001), and H\,1743-322 which has
240 and 160 Hz QPOs (Homan et al.~2005). Other examples for black hole XRBs 
are given by McClintock and Remillard (2005). Abramowicz et al.~(2003) make a case for 
ratio 2:3 frequencies in the neutron star XRB Sco X-1.

   There have been a number of attempts to understand the XRB QPOs (see reviews by van der Klis 
(2000, 2005) and Psaltis (2002)), but more recently the extension of the Psaltis-Belloni-van der Klis 
(1999) relationship (that $P_{QPO} = 15 \times P_{DNO}$) to CVs (Paper II; Mauche 2002) has 
shown that theories requiring strong gravity may not be necessary. Among them are those that enable 
both neutron and black hole XRBs to accrete into a rotating magnetosphere, as in the LIMA 
model for CVs. One example is the Robertson-Leiter model that allows proto-black holes to 
delay collapse towards their event horizons and so maintain a magnetic moment almost indefinitely 
(Robertson \& Leiter 2002, 2003). Another is the Titarchuk and Wood (2002) model that 
associates the high frequency (DNO) oscillations with the Keplerian frequency at the inner edge 
of the disc and the low frequency ones (QPOs) with magnetoacoustic oscillations that are 
excited by the transition to sub-Keplerian flow at the inner edge. Of these, the former 
more closely resembles the model that we have proposed for CVs.

   A word of caution is necessary. There are several frequencies observed in neutron
star binaries and it is only the highest of these that correlates closely with the low frequency.
There are, however, strong correlations among the other frequencies. In black hole binaries
the highest frequency analogue is not observed, but there is a correlation between other
frequencies similar to that in neutron star binaries. It is this that provides the black hole\,/\,neutron
star\,/ white dwarf general correlation (van der Klis 2005).

   Our new observations, exhibiting for the first time 3:2:1 relationships among CV DNOs, 
are a possible indication that strong gravity may not be essential for modeling DNOs and QPOs 
in compact binaries. It may also be the case that different physical processes are involved
(e.g., Klu\'zniak et al.~2005).

\section{Conclusions}

    VW Hyi continues to provide unprecedented behaviour in its rapid variations. 
No other dwarf nova is yet known that shows such a range of phenomena; whether this 
is connected with the sustained and increasing prominence of oscillations late in outburst 
is not known -- no other dwarf nova has shown this either. The rapid increase in period 
of the DNOs, at a rate greater than that observed in other declining dwarf novae, which 
we have attributed to a phase of propellering and deceleration of an equatorial accretion 
belt, has been shown here to be accompanied by the appearance of first and second harmonics 
in both DNOs and QPOs. These are again unique to VW Hyi as a dwarf nova, but bear similarities 
to harmonics seen in the QPOs of X-Ray binaries. 

   Whereas most of the behaviour of the DNOs can be understood in terms of accretion from a 
disc onto a freely rotating equatorial accretion belt, that of the QPOs has not been fully explained. 

   Tests of the models suggested in this series of papers could be made from measuring 
the amplitude and phase changes of DNOs and QPOs observed in eclipsing CVs, and from detection 
of polarization modulated at the periods of the DNOs. These will require large telescopes 
to provide the necessary good signal/noise.

\section*{Acknowledgments}

We are indebted to the late Danie Overbeek and to Rod Stubbings and Albert Jones for alerting 
us to outbursts of VW Hyi. We also thank Rod Stubbings for providing us with his long term 
light curve of VW Hyi. Retha Pretorius kindly contributed the run S7301 light curve.
BW's research is supported by the University of Cape Town; PAW is supported by research funding 
from the National Research Foundation and the University of Cape Town, and by a strategic 
grant to BW from the University of Cape Town.

\end{document}